\documentstyle[aps,prl,floats,epsf,preprint]{revtex}

\newcommand{\rmd}{{\rm d}}                                                    %
\newcommand{\rme}{{\rm e}}                                                    %
\newcommand{\rmi}{{\rm i}}                                                    %
\newcommand{\h}{{\bf h}_{\bot {\bu}}}                                         %
\newcommand{\bGamma}{{\bf\Gamma}}                                             %
\newcommand{\tsigma}{\tilde{\sigma}}                                          %
\newcommand{\tbsigma}{\tilde{\bsigma}}                                        %
\newcommand{\tn}{\tilde{n}}                                                   %
\newcommand{\tw}{\tilde{w}}                                                   %
\newcommand{\tR}{\tilde{R}}                                                   %
\newcommand{\bu}{{\bf \hat u}}                                                %
\newcommand{\br}{{\bf r}}                                                     %
\newcommand{\bsigma}{\mbox{\boldmath $\sigma$}}                               %
\newcommand{\bOmega}{{\bf\Omega}}                                             %

\begin{document}

\preprint{ }
\draft

\title{Theory of interlayer exchange coupling}

\author{P. Bruno}

\address{Max-Planck-Institut f\"ur Mikrostrukturphysik\\
Weinberg 2, D-06120 Halle, germany}

\date{\today}

\maketitle

\begin{abstract}
This paper contains the notes of lectures on 
the theory of interlayer exchange
coupling presented at the 30-th 
Ferienschule of the Institut f\"ur Festk\"orperforschung, Forschungszentrum 
J\"ulich, March 1999.
\end{abstract}

\pacs{Published in ``Magnetische Schichtsysteme'', edited by P.H.~Dederichs 
and P.~Gr\"unberg (Forschungszentrum J\"ulich, 1999).}


\section{Introduction}

Since its the discovery of interlayer exchange coupling \cite{ Parkin1990}, 
this phenomenon has stimulated a number theoretical investigations. Various 
approaches have used. These are:
\begin{itemize}
\item the Ruderman-Kittel-Kasuya-Yosida (RKKY) model \cite{ 
Yafet1987, Chappert1991, Bruno1991, Bruno1992, Coehoorn1991};
\item the free-electron model \cite{ Barnas1992, 
Erickson1993, Slonczewski1993};
\item the hole confinement model \cite{ Edwards1991, Mathon1992};
\item the Anderson 
(or $sd$-mixing) model \cite{ Wang1990, Shi1992, Bruno1992b};
\item {\em ab initio\/} calculations \cite{ Herman1991, Krompiewski1993, 
Krompiewski1994, Schilfgaarde1993, Lang1993, Nordstrom1994, Kudrnovsky1994, 
Lee1995, Kudrnovsky1996, Lang1996, Drchal1996, Stiles1996, Albuquerque1996, 
Lee1996, Mathon1997, Costa1997}.
\end{itemize}

The mechanism which now widely accepted for the IEC is based upon quantum 
intereferences in the spacer layer due to spin-dependent confinement 
\cite{ Edwards1991, Bruno1993, Bruno1995, Stiles1993}. 

These lecture notes are organized as follows. In, Section~I, the quantum 
interences due to confinement are discussed, and it is shown how this yields 
an oscillatory interlayer exchange coupling. In Section~II, the theoretical 
basis of the one-electron approach of Section~I is presented. Finally, 
in Section~III, the effect of substitutional disorder is addressed. 

\section{Physical mechanism of interlayer exchange coupling in terms of 
quantum interferences: a heuristic aproach}

The purpose of this section is to present as simply as possible the mechanism 
of interlayer exchange coupling in terms of quantum interferences due to 
electron confinement in the spacer layer. The emphasis here will be on 
physical concepts rather than on mathematical rigor. This discussion is based 
on the one given in Ref. \cite{ Bruno1995}.

\subsection{Elementary discussion of quantum confinement} 

For the sake of clarity, we shall first consider an extremely simplified 
model, namely the one-dimensional quantum well, which nevertheless contains 
the essential physics involved in the problem. Then, we shall progressively 
refine the model in order to make it more realistic.

The model consists in a one-dimensional quantum well representing the spacer 
layer (of potential $V=0$ and width $D$), sandwiched between two ``barriers'' 
$A$ and $B$ of respective widths $L_A$ and $L_B$, and respective potentials 
$V_A$ and $V_B$. Note that we use the term ``barrier'' in a general sense, 
i.e., $V_A$ and $V_B$ are not necessarily positive. Furthermore, the barrier 
widths, $L_A$ and $L_B$, can be infinite.

\subsubsection*{Change of the density of states due to quantum interferences}

Let us consider an electron of wavevector $k^+$ (with $k^+ > 0$) propagating to 
the right in the spacer layer; as this electrons arrives on barrier $B$, it 
is partially reflected to the left, with a (complex) anplitude $r_B \equiv 
|r_B| \rme^{\rmi \phi_B}$. The reflected wave of wavevector $k^-$ is in turn 
reflected on barrier $A$ with an amplitude $r_A \equiv |r_A| \rme^{\rmi 
\phi_A}$, an so on.%
\footnote[1]{\ Of course, for the one-dimensional model, one has $k^- = -k^+$; 
however, this property will generally not hold for three-dimensional 
systems to be studied below.}%
\ The module $|r_{A(B)}|$ of the reflection coefficient 
expresses the magnitude of the reflected wave, whereas the argument 
$\phi_{A(B)}$ represents the phase shift due to the reflection (note that 
the latter is not absolutely determined and depends on the choice of the 
coodinate origin).

\begin{figure}
\includegraphics{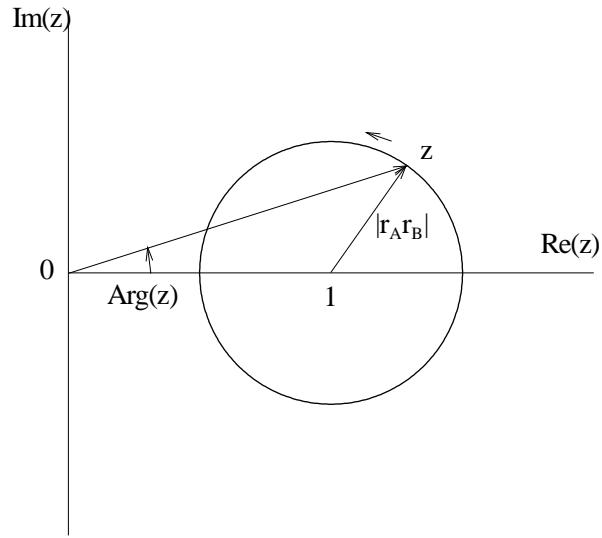}
\vspace*{9cm}
\caption{Graphical interpretation of equation (\protect\ref{eq:th:DeltaN}).}
\label{fig:graphic}
\end{figure}

The interferences between the waves due to the multiple reflections on the 
barriers induce a modification of the density of states in the spacer layer, 
for the electronic state under consideration. The phase shift resulting from 
a complete round trip in the spacer is
\begin{equation}
\Delta \phi = q D + \phi_A + \phi_B \ ,
\end{equation}
with 
\begin{equation}
q \equiv k^+-k^- \ .
\end{equation}
If the interferences are constructive, i.e., if
\begin{equation}
\Delta \phi = 2n\pi
\end{equation}
with $n$ an integer, one has an increase of the density of states; conversely, 
if the interferences are destructive, i.e., if
\begin{equation}
\Delta\phi = (2n+1)\pi 
\end{equation}
one has a reduction of the density of states. Thus, in a first approximation, 
we expect the modification of the density of states in the spacer, 
$\Delta n(\varepsilon )$, to vary with $D$ like
\begin{equation}
\cos \left( q D + \phi_A +\phi_B \right) .
\end{equation}
Furthermore, we expect that this effect will proportional to the amplitude 
of the reflections on barriers $A$ and $B$, i.e., to $|r_A r_B|$; finally, 
$\Delta n(\varepsilon )$ must be proportional to the width $D$ of the spacer 
and to the density of states per unit energy and unit width,
\begin{equation}
\frac{2}{\pi} \frac{\rmd q }{\rmd\varepsilon}
\end{equation}
which includes a factor of $2$ for spin degenaracy. We can also include the 
effect of higher order interferences due to $n$ round trips in the spacer; the 
phase shift $\Delta \phi$ is then multiplied by $n$ and $|r_A r_B|$ is 
replaced by $|r_A r_B|^n$. Gathering all the terms, we get,
\begin{eqnarray}
\Delta n(\varepsilon ) &\approx& \frac{2D}{\pi}\, 
\frac{\rmd q}{\rmd\varepsilon} \sum_{n=1}^\infty \, |r_Ar_B|^n \, 
\cos n\left( qD +\phi_A +\phi_B \right) \nonumber \\
&=& \frac{2}{\pi} \mbox{ Im} \left( \rmi D\, 
\frac{\rmd q}{\rmd\varepsilon}\, \sum_{n=1}^\infty \, 
\left(r_Ar_B\right)^n \, \rme^{n \rmi qD} \right) \nonumber \\
&=& \frac{2}{\pi} \mbox{ Im} \left( \rmi \, 
\frac{\rmd q}{\rmd\varepsilon}\, \frac{r_Ar_B\, 
\rme^{\rmi qD}}{1 - r_Ar_B\, 
\rme^{\rmi qD}} \right)
\end{eqnarray}
As will appear clealy below, it is more convenient to consider the integrated 
density of states
\begin{equation}
N(\varepsilon ) \equiv \int_{-\infty}^{\varepsilon} n(\varepsilon^\prime )\, 
\rmd\varepsilon^\prime  .
\end{equation}
The modification $\Delta N (\varepsilon)$ of the intergated density of states 
due to electron confinement is
\begin{eqnarray}\label{eq:th:DeltaN}
\Delta N(\varepsilon ) &=& \frac{2}{\pi} \mbox{ Im}\, \sum_{n=1}^\infty \, 
\frac{\left( r_A r_B \right)^n}{n}\, \rme^{n\rmi qD} \nonumber \\
&=& -\, \frac{2}{\pi} \mbox{ Im } \ln \left( 1 -r_A r_B\, \rme^{\rmi qD} 
\right)
\end{eqnarray}
A simple graphical interpretation of the above expression can be obtained by 
noting that $\mbox{Im }\ln (z) = \mbox{Arg }(z)$, for $z$ complex; thus, 
$\Delta N(\varepsilon )$ is given by the argument, in complex plane, of a 
point located at an angle $\Delta \phi = qD + \phi_A + \phi_B$ on a 
circle of 
radius $|r_A r_B|$ centred in 1. This graphical construction is shown in 
\ref{fig:graphic}.

\begin{figure}
\includegraphics{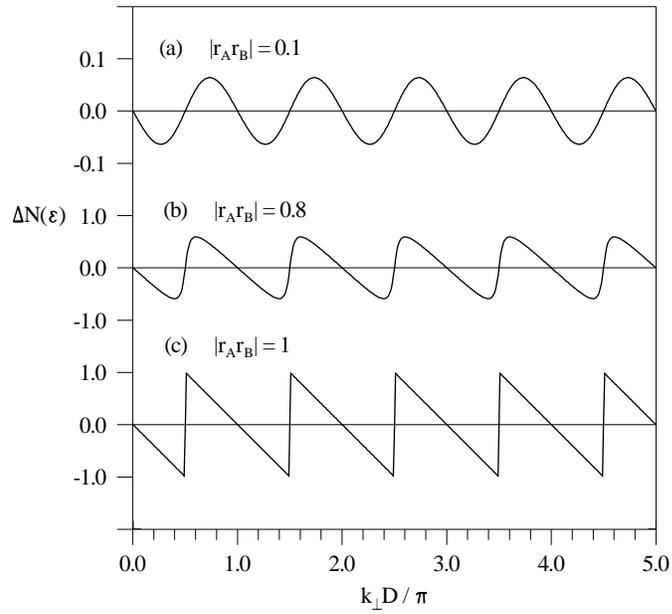}
\vspace*{10cm}
\caption{Variation of $\Delta N(\varepsilon )$ as a function of $D$, for 
various values of the confinement strength: (a) $|r_A r_B| = 0.1$, 
(b) $|r_A r_B| = 0.8$, (c) $|r_A r_B| = 1$ (full confinement). Note the 
different scales along the ordinate axis.}
\label{fig:deltaN}
\end{figure}

The variation of $\Delta N(\varepsilon )$ as a function of $D$ is shown in 
\ref{fig:deltaN}, for various values of the confinement strength $|r_A r_B|$. 
For weak confinement (a), $\Delta N(\varepsilon )$ varies with $D$ in 
sinusoidal manner. As one increases the confinement strength (b), the 
oscillations are distorded, due to higher order interferences. Finally, for 
full confinement (c), $\Delta N(\varepsilon )$ exhibits some jumps that 
correspond to the bound states. We note however, that the period $\Lambda$ of 
the oscillations of $\Delta N(\varepsilon)$ does not depend on the confinement 
strength, but only on the wavevector $q \equiv k^+-k^-$, namely, 
$\Lambda = 2\pi /q$.

So far, we have implicitely restricted ourselves to positive energy states. 
Negative energy states (i.e., of imaginary wavevector) are forbidden in 
absence of the barriers $A$ and $B$, because their amplitude diverges either 
on the right hand side or on the left hand side, so that they cannot be 
normalized. This matter of fact no longer holds in the presence of the 
barriers if $V_A$ (or $V_B$ or both) is negative: the negative energy states, 
i.e., varying exponentially in the spacer, can be connected to allowed states 
of $A$ or $B$. In order to treat these states consistently, we simply have to 
extend the concept of reflection coefficient to to states of imaginary 
wavevector, which is straightforward. One can check that, with this 
generalization, \ref{eq:th:DeltaN} acounts properly for the contribution of 
the evanescent states. Physically, this can be interpretated as a coupling of 
$A$ and $B$ by tunnel effect. 

\subsubsection*{Energy associated with the quantum interferences in the 
spacer}

Let us now study the modification of the energy of the system due to the 
quantum interferences. In order to conserve the total number of electron, 
it is convenient to work within the grand-canonical ensemble, and to consider 
the thermodynamic grand-potential, which is given by
\begin{eqnarray}
\Phi &\equiv& - k_BT \int_{-\infty}^{+\infty} \ln \left[ 1+ \exp 
\left( \frac{\varepsilon_F - \varepsilon}{k_BT}\right) \right] \, 
n(\varepsilon )\, \rmd\varepsilon \nonumber \\
&=& - \int_{-\infty}^{+\infty} N(\varepsilon )\, f(\varepsilon)\, 
\rmd\varepsilon .
\end{eqnarray}
At $T=0$, this reduces to 
\begin{eqnarray}
\Phi &\equiv& \int_{-\infty}^{\varepsilon_F} \left( \varepsilon - 
\varepsilon_F \right)\, n(\varepsilon )\, \rmd\varepsilon \nonumber \\
&=& -\int_{-\infty}^{\varepsilon_F} N(\varepsilon)\, \rmd\varepsilon .  
\end{eqnarray}
The energy $\Delta E$ associated with the interferences is the contribution 
to $\Phi$ corresponding to $\Delta N(\varepsilon )$, 
\begin{equation}
\Delta E = \frac{2}{\pi} \mbox{ Im} \int_{-\infty}^{+\infty} \ln 
\left( 1 - r_Ar_B \, \rme^{\rmi qD} \right) \, \rmd\varepsilon .
\end{equation}

\subsubsection*{Three-dimensional layered system}

The generalization of the above discussion to the more realistic case of a 
three-dimensional layered system is immediate. Since the system is 
invariant by translation parallely to the plane, so that the in-plane 
wavevector $\bf{k}_\|$ is a good quantum number. Thus, for a given 
$\bf{k}_\|$, one has an effective one-dimensional problem analogous to the 
one discussed above. The resulting effect of quantum intereferences is 
obtained by summing over $\bf{k}_\|$. The modification of integrated density 
of states per unit area is
\begin{equation}
\Delta N(\varepsilon ) = -\,\frac{1}{2\pi^3}\mbox{ Im} \int
\rmd^2\mbox{\bf k}_\|  \, \ln\left( 1- r_Ar_B\, \rme^{\rmi q_\bot D} \right) \ ,
\end{equation}
and the coupling energy per unit area is 
\begin{equation}\label{eq:th:QSE-Energy-3D}
\Delta E = \frac{1}{2\pi^3}\mbox{ Im} \int \rmd^2\mbox{\bf k}_\|
\int_{-\infty}^{+\infty}f(\varepsilon )\, \ln\left( 1- r_Ar_B\, 
\rme^{\rmi q_\bot D}
\right) \, \rmd\varepsilon \ .
\end{equation}

\subsubsection*{Quantum size effect in an overlayer}

The case of a thin overlayer deposited on a substrate is of considerable 
interest. In this case, one of the barriers (say, $A$) consists of the vacuum, 
and barrier $B$ is constituted by the substrate itself. The potential of the 
vacuum barrier is $V_{\rm vac} = \varepsilon_F + W$, where $W$ is the the work 
function; thus it is perfectly reflecting for occupied states, i.e., 
$|r_{\rm vac} + 1|$. On the other hand, the reflection on the substrate (or 
coefficent $r_{\rm sub}$) may be total or partial, depending on the state 
under consideration.

The spectral density of the occupied states in the overlayer can be 
investigated experimentally by photoemission spectroscopy; in addition, by 
using inverse photoemission, one can study the unoccupied states. If 
furthermore these techniques are used in the ``angle-resolved'' mode, they 
give information on the spectral density \emph{locally in the $\bf{k}_\|$ 
plane}.

For a given thickness of the overlayer, the photoemission spectra (either 
direct or inverse) exhibit some maxima and minima corresponding, respectively, 
to the energies for which the interferences are constructive and destructive. 
When the confinement is total, narrow peaks can be observed, which correspond 
to the quantized confined states in the overlayer, as was pointed out by Loly 
and Pendry \cite{ Loly1983}.

Quantum size effects due to electron confinement in the photoemission spectra 
of overlayers have been observed in various non-magnetic systems 
\cite{ Wachs1986, Lindgren1987, Lindgren1988, Lindgren1988b, Lindgren1989, 
Miller1988, Mueller1989, Mueller1990, Jalochowski1992} In particular, the 
systems Au(111)/Ag/vacuum and Cu(111)/Ag/vacuum offer excellent examples of 
this phenomenon \cite{ Miller1988, Mueller1990}.

\subsubsection*{Paramagnetic overlayer on a ferromagnetic substrate: 
Spin-polarized quantum size effect} 

So far our discussion concerned exclusively non-magnetic systems. 
Qualitatively new behavior can be expected some of the layers are 
ferromagnetic.  case of particular interest is the one of a paramagnetic 
overlayer on a ferromagnetic substrate. 

In the interior of the overlayer, the potential is independent of the spin; 
therefore the propagation of electrons is described by a wave vector $k_\bot$ 
which is spin-independent. The reflection coefficient on the vacuum barrier, 
$r_{\rm vac}$, is also spin-independent. However, the ferromagnetic substrate 
constitutes a spin-dependent potential barrier; thus, the substrate reflection 
coefficients for electrons with a spin parallel to the majority and minority 
spin directions of the substrate, respectively $r^\uparrow_{\rm sub}$ and 
$r^\downarrow_{\rm sub}$. It is convenient to define the spin average 
\begin{equation}
\overline{r}_{\rm sub} \equiv \frac{r_{\rm sub}s^\uparrow + 
r_{\rm sub}^\downarrow}{2}
\end{equation}
and the spin asymmetry 
\begin{equation}
\Delta r_{\rm sub} \equiv \frac{r_{\rm sub}^\uparrow - 
r_{\rm sub}^\downarrow}{2} .
\end{equation}
In this case, the electron confinement in the overlayer gives rise to a 
spin-dependent modulation of the spectral density versus overlayer thickness; 
the period of the modulation is the same for both spins, whereas the amplitude 
and phase are expected to be spin-dependent. 

The quantum size effects in paramagnetic overlayers on a ferromagnetic 
substrate have been investigated by several groups \cite{ Brookes1991, 
Ortega1992, Ortega1993, Ortega1993b, Garrison1993, Carbone1993, Smith1994, 
Johnson1994, Himpsel1995, Crampin1996, Segovia1996, Klaesges1998}. The 
systems studied most are Cu overlayers on a Co(001) substrate and Ag 
overlayers on a Fe(001) substrate. 
Ortega and Himpsel \cite{ Ortega1992, Ortega1993} observed a quantum size 
effect in the normal-emission photoelectron spectra of copper overlayer on 
fcc cobalt (001) substrate. They observed peaks due to quantum size effects 
both in the photoemission and in the inverse photoemission spectra an 
oscillation of the photoemission intensity. These quantum size effects 
manifest themselves also by an oscillatory behavior of the photoemission 
intensity at the Fermi level; as the observed oscillation period (5.9~atomic 
layers) is close to the long period of interlayer exchange coupling 
oscillations in Co/Cu(001)/Co, they pointed out that the two phenomena are 
related to each other; they also claimed that the observed oscillations in 
photoemission are spin dependent and due mostly to minority electrons. A 
direct confirmation of this conjecture has been given independently by 
Garrison {\em et al.\/} \cite{ Garrison1993} and by Carbone {\em et al.\/} 
\cite{ Carbone1993} 
by means of spin-polarized photoemission. They found that both the 
intensity and the spin-polarization exhibit an oscillatory behavior with the 
same period (5 -- 6 atomic layers), but they have opposit phases, which 
indicates that the quantum size effect does indeed take place predominantly 
in the minority-spin band as proposed by Ortega and Himpsel \cite{ Ortega1992, 
Ortega1993}. 
Recently, Kl\"asges {\em et al.\/} \cite{ Klaesges1998} have observed 
spin-polarized 
quantum size effects in a copper overlayer on cobalt (001) for a non-zero 
in-plane wave vector corresponding to the short period oscillation of 
interlayer exchange coupling in Co/Cu(001)/Co; they observed short period 
oscillations of the photoemission intensity in good agrement with the short 
period oscillations of interlayer coupling. This observation provides a 
further confirmation of the relation between quantum size effects in 
photoemission and oscillation of interlayer exchange coupling. 

Photoemission studies of quantum size effects have also been performed in 
other kinds of systems such as ferromagnetic overlayer on a non-magnetic 
substrate, or systems comprising more layers \cite{ Himpsel1991, Ortega1993c, 
Li1995, Himpsel1995b, Li1997}. 

Photoemission spectroscopy undoubtly constitutes a method of choice for 
investigating quantum size effects in metallic overlayers: this is due to its 
unique features, which allow selectivity in energy, in-plane wave vector, and 
spin. 

Besides photemission, spin-polarized quantum size effects in paramagnetic 
overlayers on a ferromagnetic substrate are also responsible for oscillatory 
behavior (versus overlayer thickness) of spin-polarized secondary electron 
emission \cite{ Koike1994, Furukawa1996}, linear \cite{ Bennett1990, 
Katayama1993, Carl1995, Megy1995, Bruno1996c, Suzuki1998} and non-linear 
\cite{ Luce1996, Kirilyuk1996} magneto-optical Kerr, and magnetic anisotropy 
\cite{ Weber1996, Back1997}. However, these effects usually involve a 
summation over all electronic states, so that the quantitative analysis of 
the quantum size effects may be fairly complicated.

\subsection{Interlayer exchange coupling due to quantum interferences}

Let us now consider the case of a paramagnetic layer sandwiched between two 
ferromagnetic barriers $A$ and $B$. Now, the reflection coefficients on both 
sides of the paramagnetic spacer layer are spin dependent. \emph{A priori} the 
angle $\theta$ between the magnetizations of the two ferromagnetic can take 
any value; however, for the sake of simplicity, we shall restrict ourselves 
here to the ferromagnetic (F) configuration (ie., $\theta =0$) and the 
antiferromagnetic (AF) one (i.e., $\theta = \pi$).  

For the ferromagnetic configuration, the energy per unit area due to quantum 
interference is easily obtained from \ref{eq:th:QSE-Energy-3D}, i.e.,
\begin{equation}
\Delta E_F = \frac{1}{4\pi^3}\mbox{ Im} \int \rmd^2 {\bf k}_\|
\int_{-\infty}^{+\infty} f(\varepsilon )
\, \left[ \ln\left( 1- r_A^\uparrow r_B^\uparrow
\rme^{\rmi q_\bot D} \right) 
+ \ln\left( 1- r_A^\downarrow
r_B^\downarrow \rme^{\rmi q_\bot D} \right) \right] \, \rmd\varepsilon \ .
\end{equation}
In this equation, the first and the second term correspond respectively to 
majority- and  minority-spin electrons. The antiferromagnetic conguration is 
obtained by reversing the magnetization of $B$, i.e., by interchanging 
$r_B^\uparrow$ and $r_B^\downarrow$; thus the corresponding energy per unit 
area is
\begin{equation}
\Delta E_{AF} = \frac{1}{4\pi^3}\mbox{ Im} \int \rmd^2 {\bf k}_\|
\int_{-\infty}^{+\infty} f(\varepsilon )
\, \left[ \ln\left( 1- r_A^\uparrow r_B^\downarrow
\rme^{\rmi q_\bot D} \right) 
+ \ln\left( 1- r_A^\downarrow
r_B^\uparrow \rme^{\rmi q_\bot D} \right) \right] \, \rmd\varepsilon \ .
\end{equation}
Thus, the interlayer exchange coupling energy is
\begin{equation}
E_F - E_{AF} = \frac{1}{4\pi^3}\mbox{ Im} \int \rmd^2 {\bf k}_\|
\int_{-\infty}^{+\infty}f(\varepsilon )\, 
\ln\left[ \frac{ \left( 1-
r_A^\uparrow r_B^\uparrow \rme^{\rmi q_\bot D} \right)\left( 1-
r_A^\downarrow r_B^\downarrow \rme^{\rmi q_\bot D} \right)}{ \left( 1-
r_A^\uparrow r_B^\downarrow \rme^{\rmi q_\bot D} \right)\left( 1-
r_A^\downarrow r_B^\uparrow \rme^{\rmi q_\bot D} \right)}\right] \,
\rmd\varepsilon 
\end{equation}
which can be simplified as 
\begin{equation}
E_F - E_{AF} \approx -\ \frac{1}{\pi^3}\mbox{ Im} \int \rmd^2 {\bf k}_\|
\int_{-\infty}^{\infty}f(\varepsilon )\, \Delta r_A \Delta r_B\,
\rme^{\rmi q_\bot D}  \, \rmd\varepsilon 
\end{equation}
in the limit of weak confinement. The above expression for the IEC has a rather 
transparent physical interpretation. First, as the integrations on 
{\bf k}$_\|$ over the first two-dimensional Brillouin zone and on the energy up 
to the Fermi level show, the IEC is a sum of contributions from all occupied 
electronic states. The contribution of a given electronic state, of energy 
$\varepsilon$ and in-plane wavevector {\bf k}$_\|$, consists of the product 
of three factors: the two factors $\Delta r_A$ and $\Delta r_B$ express the 
spin-asymmetry of the confinement due to the magnetic layers $A$ and $B$, 
respectively, while 
the exponential factor $e^{iq_\bot D}$ describes the propagation 
through the spacer and is responsible for the interference (or quantum size) 
effect. Thus, this approach establishes an explicit and direct link between 
oscillatory IEC and quantum size effects such as observed in photoemission.

\subsubsection*{Asymptotic behavior for large spacer thicknesses}

In the limit of large spacer thickness $D$, the exponential factor oscillates 
rapidly with $\varepsilon$ and {\bf k}$_\|$, which leads to some cancellation 
of the contributions to the IEC due to the different electronic states. 
However, because the integration over energy is abruptly stopped at 
$\varepsilon_F$, states located at the Fermi level give predominant 
contributions. Thus the integral on $\varepsilon$ may be calculated
by fixing all other factors to their value at $\varepsilon_F$,
and by developing $q_\bot \equiv k_\bot^+ -k_\bot^-$ around
$\varepsilon_F$, i.e.,
\begin{equation}
q_\bot \approx q_{\bot F} +2 \frac{\varepsilon -\varepsilon_F}
{\hbar v_{\bot F}^{+-}} ,
\end{equation}
with
\begin{equation}
\frac{2}{v_{\bot F}^{+-}} \equiv \frac{1}{v_{\bot F}^+} -
\frac{1}{v_{\bot F}^-} .
\end{equation}
The integration (see Ref.~\cite{ Bruno1995} for details) yields
\begin{eqnarray}
E_F-E_{AF} &=& \frac{1}{2\pi^3} \mbox{ Im} \int d^2\mbox{\bf k}_\|\,
\frac{i\hbar v_{\bot F}^{+-}}{D}\, \Delta r_A\Delta r_B
e^{iq_{\bot F}D} \nonumber \\
&&\times F(2\pi\, k_BT\,D/\hbar v_{\bot F}^{+-}) ,
\end{eqnarray}
where
\begin{equation}
F(x) \equiv \frac{ x}{\sinh x} .
\end{equation}
In the above equations, $q_{\bot F}$ is a vector spanning the
{\em complex Fermi surface\/}; the velocity $v_{\bot F}^{+-}$ is
a combination of the group velocities at the
extremities $k_{\bot F}^+$ and $k_{\bot F}^-$.

Next, the integration on $\mbox{\bf k}_\|$ is performed by noting,
that, for large spacer thickness $D$, the only significant
contributions arise from the neighboring of critical vectors
$\mbox{\bf k}_\|^\alpha$ where $q_{\bot F}$ is stationary. Around such
vectors, $q_{\bot F}$ may be expanded as
\begin{equation}
q_{\bot F} = q_{\bot F}^\alpha - \frac{ \left( k_x-k_x^\alpha
\right)^2} {\kappa_x^\alpha} - \frac{ \left( k_y-k_y^\alpha
\right)^2} {\kappa_y^\alpha} ,
\end{equation}
where the cross terms have been canceled by a proper choice of
the axes; $\kappa_x^\alpha$ and $\kappa_y^\alpha$
are combinations of the curvature radii of the Fermi
surface at $(\mbox{\bf k}_\|^\alpha ,k_{\bot}^{+\alpha})$ and 
$(\mbox{\bf
k}_\|^\alpha ,k_{\bot}^{-\alpha})$. 

The integral is calculated by using the stationary phase
approximation,\cite{ Bruno1995} and one obtains
\begin{eqnarray}
J_1 &=& \mbox{ Im } \sum_\alpha \frac{ \hbar v_\bot^\alpha
\kappa_\alpha}{4\pi^2D^2} \Delta r_A^\alpha \Delta r_B^\alpha
e^{iq_\bot^\alpha D}\nonumber \\
&&\times F(2\pi k_BTD/\hbar v_\bot^\alpha ) ,
\end{eqnarray}
where $q_\bot^\alpha$, $v_\bot^\alpha$, $\Delta r_A^\alpha$,
$\Delta r_B^\alpha$ correspond to the critical vector $\mbox{\bf
k}_\|^\alpha$, and
\begin{equation}
\kappa_\alpha \equiv  \left(\kappa_x^\alpha\right)^{1/2}
\left(\kappa_y^\alpha \right)^{1/2} ;
\end{equation}
in the above equation, one takes the square root with an argument
between $0$ and $\pi$.

\begin{figure}[ht] 
\includegraphics{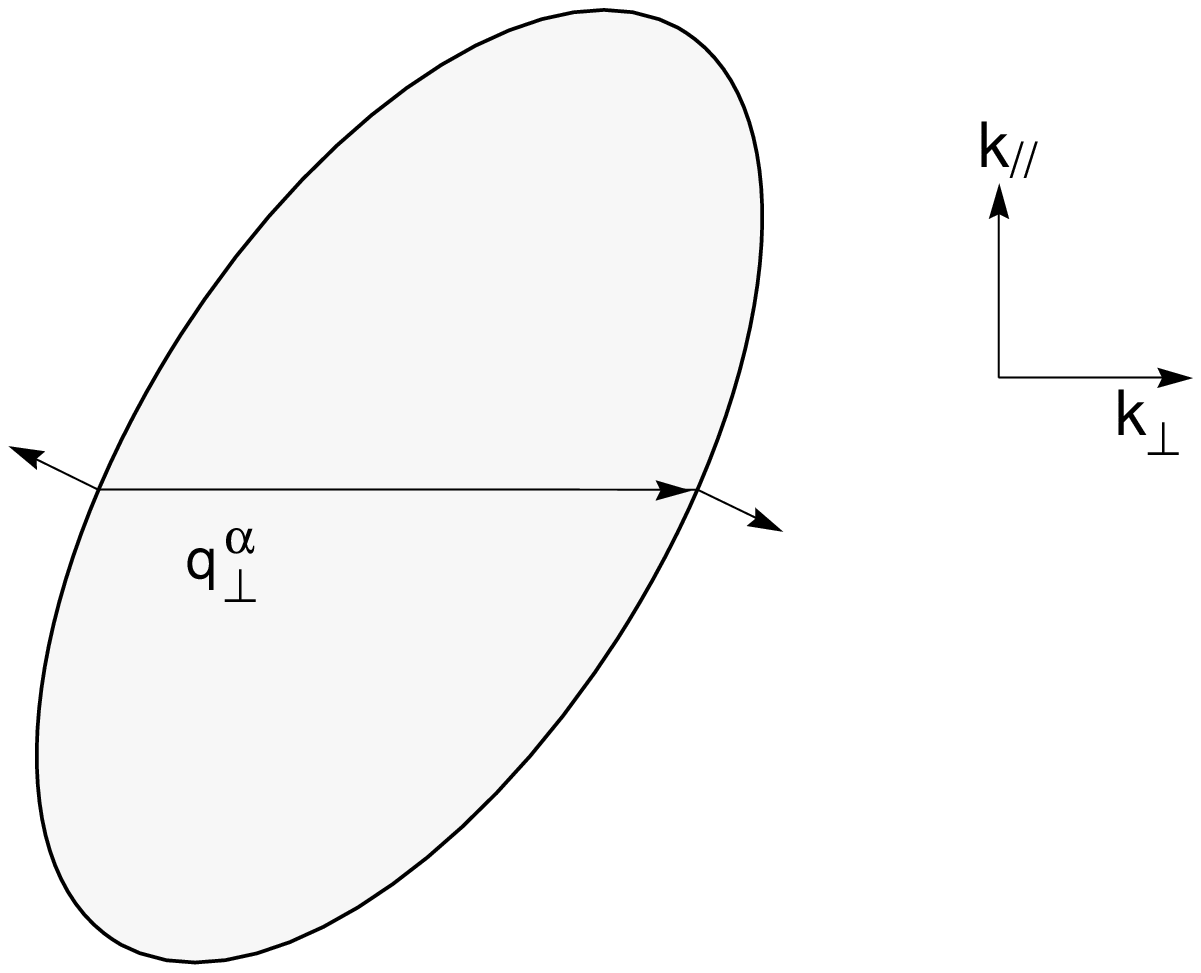}
\vspace*{8cm}
\caption{Sketch showing the wavevector $q_\bot^\alpha$ giving the oscillation
period of oscillatory interlayer exchange coupling, for the case of a 
non-spherical Fermi 
surface.}
\label{ fig_spanning}
\end{figure}

This analysis shows 
that {\em in fine\/}, the only remaining terms in the limit of large spacer 
thickness $D$ arise from the neighborhood of states having in-plane wavevectors 
{\bf k}$_\|^\alpha$ such that the spanning vector of the Fermi surface 
$q_{\bot F} = k_{\bot F}^+ - k_{\bot F}^-$ is stationary with respect to 
{\bf k}$_\|$ for {\bf k}$_\| =${\bf k}$_\|^\alpha$, and the corresponding 
contribution oscillates with a wavevector equal to $q_{\bot F}^\alpha$. This 
selection rule was first derived in the context of the RKKY model 
\cite{ Bruno1991}; it is illustrated in Fig.~\ref{ fig_spanning}. 
There may be several such stationary spanning vectors and, hence, 
several oscillatory components; they are labelled by the index $\alpha$.

\begin{figure}[h]
\includegraphics{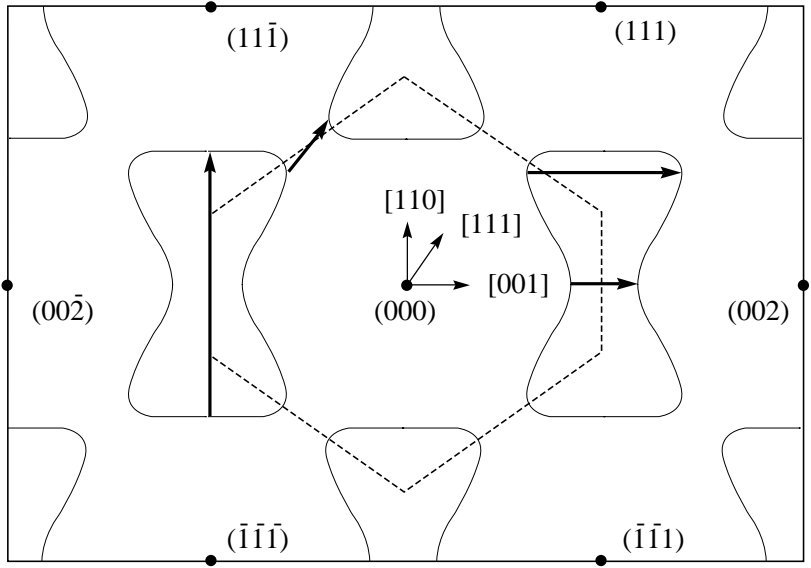}
\vspace*{9cm}
\caption{Cross section of the Fermi surface of Cu along the $(1\bar 10)$ plane 
passing through the origin. The solid dots indicate the reciprocal alttice 
vectors. The dashed lines indicate the boundary of the first brilouin zone. 
The solid arrows, respectiveley horizontal, oblique, and vertical, indicate the
vectors $q_\bot^\alpha$ giving the oscillation period(s), respectively for the 
$(001)$, $(111)$, and $(110)$ orientations.}
\label{ fig_Fermi_Cu}
\end{figure}

\begin{table}
\caption{Comparison between the theoretical predictions of 
Ref.~\protect\cite{ Bruno1991}and 
experimental observations for the oscillation periods of interlayer 
exchange coupling versus overlayer thickness.}
\label{ tab_period}
\vspace*{0.5\baselineskip}
\begin{tabular}{ccccc}
spacer & theoretical periods & system & experimental periods & Ref. 
\\
\tableline
        &                        & Co/Cu/Co(111) & $\Lambda \approx$\dec 5. AL 
& \cite{ Parkin1991} \\
\cline{3-5}
Cu(111) & $\Lambda =$\dec 4.5 AL & Co/Cu/Co(111) & $\Lambda \approx$\dec 
6. AL & \cite{ Mosca1991}  \\
\cline{3-5}
        &                        & Fe/Cu/Fe(111) & $\Lambda \approx$\dec
6. AL & \cite{ Petroff1991} \\
\tableline
        &                        & Co/Cu/Co(001) & $\Lambda \approx$\dec 6. AL 
& \cite{ Miguel1991} \\
\cline{3-5}
Cu(001) & $\Lambda_1 =$\dec 2.6 AL & Co/Cu/Co(001) & $\Lambda_1 \approx$\dec 
2.6 AL & \cite{ Johnson1992}  \\
        & $\Lambda_2 =$\dec 5.9 AL &               & $\Lambda_2 \approx$\dec 
8. AL &                       \\
\cline{3-5}
        &                        & Co/Cu/Co(001) & $\Lambda_1 \approx$\dec 
2.7 AL & \cite{ Weber1995} \\
        &                        &               & $\Lambda_2 \approx$\dec
6.1 AL \\
\cline{3-5}
        &                        & Fe/Cu/Fe(001) & $\Lambda \approx$\dec 7.5 AL 
& \cite{ Bennett1990} \\
\tableline
Ag(001) & $\Lambda_1 =$\dec 2.4 AL & Fe/Ag/Fe(001) & $\Lambda_1 \approx$\dec 
2.4 AL & \cite{ Unguris1993}  \\
        & $\Lambda_2 =$\dec 5.6 AL &               & $\Lambda_2 \approx$\dec 
5.6 AL &                       \\
\tableline
Au(001) & $\Lambda_1 =$\dec 2.5 AL & Fe/Au/Fe(001) & $\Lambda_1 \approx$\dec 
2. AL & \cite{ Fuss1992}  \\
        & $\Lambda_2 =$\dec 8.6 AL &               & $\Lambda_2 \approx$\dec  
7. $\!\!\!\!\!\!\!\!\!\! -8$ AL &                       \\
\cline{3-5}
       &                           & Fe/Au/Fe(001) & $\Lambda_1 \approx$\dec 
2.5 AL & \cite{ Unguris1994, Unguris1997}  \\
        &                          &               & $\Lambda_2 \approx$\dec 
8.6 AL &                       \\
\end{tabular}
\end{table}

The above selection rule allows to predict the oscillation period(s) of the 
interlayer exchange coupling versus spacer thickness by just inspecting the 
bulk Fermi surface of the spacer material. In view of an experimental test of 
these predictions, noble metal spacer layers appear to be the best suited 
candidates; there are several reasons for this choice:
\begin{itemize}
\item Fermi surfaces of noble metals are known very accurately from de Haas-van 
Alphen and cyclotron resonance experiments \cite{ Halse1969};
\item since only the $sp$ band intersect the Fermi level, the Fermi surface is 
rather simple, and does not depart very much from a free-electron Fermi sphere;
\item samples of very good quality with noble metals as a spacer layer could be 
prepared.
\end{itemize}

Fig.~\ref{ fig_Fermi_Cu} shows a cross-section of the Fermi surface of Cu, 
indicating the stationary spanning vectors for the (001), (111), and (110) 
crystalline orientations \cite{ Bruno1991}; the Fermi surfaces of Ag and Au are 
qualitatively similar. For the (111) orientation, a single (long) period is 
predicted; for the (001) orientation, both a long period and a short period are 
predicted; for the (110) orientation, four different periods are predicted 
(only one stationary spanning vector is seen in figure \ref{ fig_Fermi_Cu}, 
the three others being located in other cross-sections of the Fermi surface). 
These theoretical predictions have been confirmed successfully by numerous 
experimental observations. In particular, the coexistence of a long and a short 
period for the (001) orientation has been confirmed for Cu \cite{ Johnson1992, 
Weber1995}, 
Ag \cite{ Unguris1993}, and Au \cite{ Fuss1992, Unguris1994, Unguris1997}; 
and the experimental periods have been found to be in excellent agreement 
with the theoretical ones. 

In a further attempt to test the theoretical predictions for the periods of 
oscillatory coupling, several groups \cite{ Okuno1993, Parkin1993, Bobo1993}
have undertaken to modify in a controlled manner the size of the Fermi surface 
(and hence, the period of the coupling) by alloying the spacer noble metal (Cu) 
with a metal of lower valence (Ni); in both cases, the change in oscillation 
period due to alloying has been found in good agreement with the expected 
change in the Fermi surface. 

Although the asymptotic approximation is often an excellent one to describe 
the IEC, in cases where the reflection coefficients vary strongly with 
$\varepsilon$ and ${\bf k}_\|$ near the stationary spanning vectors 
$q_\bot^\alpha$ of the spacer Fermi surface, preasymptotic corrections nead 
to be considered. A detailled discussion of the preasymptotic corrections is  
presented in Ref.~\cite{ Bruno1998}.

\subsubsection*{Effect of magnetic layer thickness}

\begin{figure}
\includegraphics{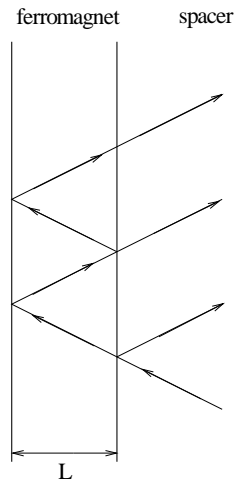}
\vspace*{7cm}
\caption{Sketch of the waves contributing to the net reflection
coefficient on a ferromagnetic layer of finite thickness $L$.}
\label{ fig_multi_refl}
\end{figure}

\begin{figure}
\includegraphics{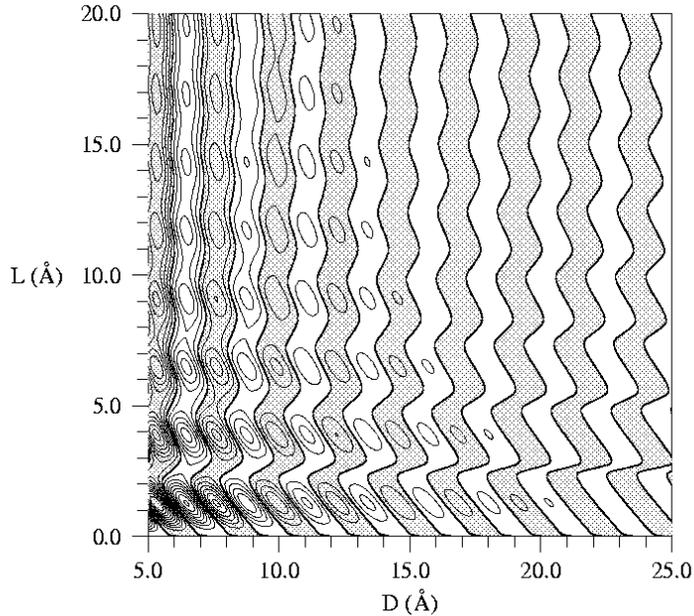}
\vspace*{11.5cm}
\caption{Contour plot of the interlayer exchange coupling constant
$J\equiv (E_F-E_{AF})/2$
vs spacer thickness $D$ and magnetic layer thickness
$L$, calculated within the free-electron model (see 
Ref.~\protect\cite{ Bruno1995} for details). The
spacing between successive contour lines is $40\times 10^{-3}$
erg~cm${}^{-2}$; the shaded area corresponds to antiferromagnetic
coupling.}
\label{ fig_ferro_thick}
\end{figure}

As already mentioned, the influence of the IEC on the ferromagnetic layer 
thickness is contained in the reflection coefficients $\Delta r_A$ and 
$\Delta r_B$. If the ferromagnetic layers are of finite thickness, reflections 
usually may take place at the two interfaces bounding the ferromagnetic layers, 
as sketched in Fig.~\ref{ fig_multi_refl},
giving rise to interferences \cite{ Bruno1993b}, and hence, to oscillations of 
the IEC versus ferromagnetic layers thickness. A more detailled discussion of 
this effect is given in Refs. \cite{ Bruno1995, Bruno1993b}. This behavior was 
first predicted from calculations based upon a free-electron model 
\cite{ Barnas1992}. 
On the experimental point of view, it 
was confirmed by Bloemen {\em et al.\/} \cite{ Bloemen1994} in Co/Cu/Co(001) and 
by Okuno and Inomata \cite{ Okuno1994} in Fe/Cr/Fe(001). The amplitude of the 
oscillations of the IEC versus ferromagnetic layers thickness is generally much 
smaller than the oscillations versus spacer thickness, and do not give rise to 
changes of sign of the IEC. This is illustrated in Fig.~\ref{ fig_ferro_thick}.

\subsubsection*{Effect of overlayer thickness}

A more ({\em a priori\/}) surprising behavior is the dependence of the IEC on 
the thickness of the protective overlayer. From a na\"\i ve point of view, one 
might think that layers external to the basic ferromagnet/spacer/ferromagnet 
sandwich should not influence the interaction between the two ferromagnetic 
layers. This view is incorrect, in particular when the system is covered by an 
ultrathin protective overlayer. In this case, the electrons are able to reach 
the vacuum barrier, which is a perfectly reflecting one, so that strong 
confinement and interference effects take place in the overlayer, which lead to 
a weak but sizeable oscillatory variation of the IEC as a function of the 
overlayer thickness. 

\begin{figure}
\includegraphics{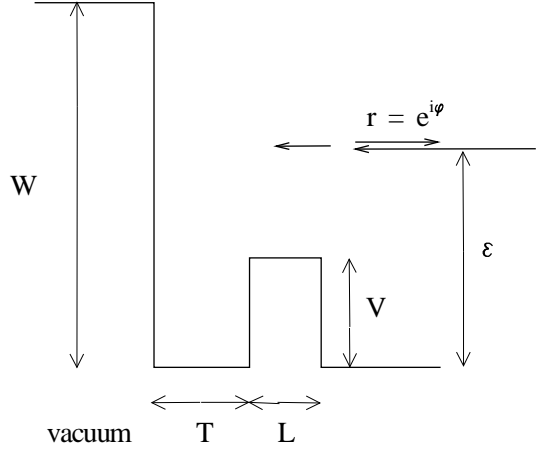}
\vspace*{9cm}
\caption{Sketch of the model used to discuss the influence of the overlayer 
and vacuum barrier on the reflection coefficient.}
\label{ fig_model1}
\end{figure}

\begin{figure}
\includegraphics{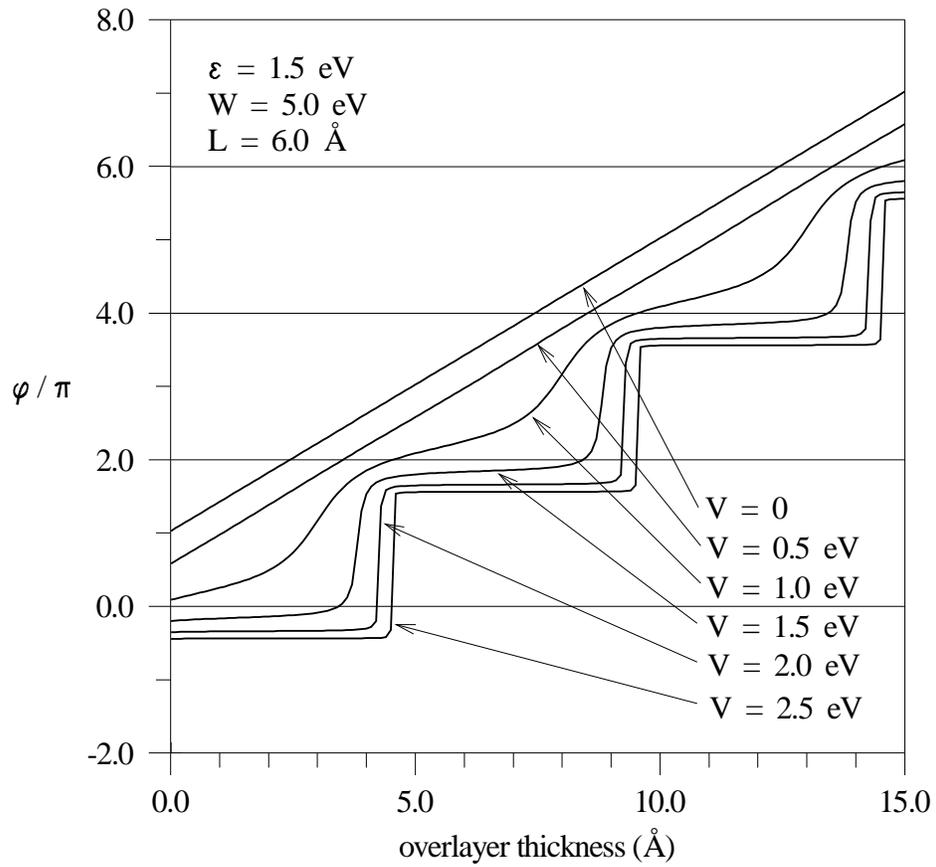}
\vspace*{12.5cm}
\caption{Variation of the reflection phase shift $\phi$ versus overlayer 
thickness $T$ for various values of the height $V$ of the magnetic barrier, as 
indicated by the arrows. Parameters: $\varepsilon = 1.5$~eV, $W=5.0$~eV, 
$L=6.0$~\AA .}
\label{ fig_refl}
\end{figure}

Here, we illustrate the effect of interferences in the overlayer by means of 
the simple model depicted in Fig.~\ref{ fig_model1}. An electron of energy 
$\varepsilon$ and in-plane wave 
vector {bf k}${}_\| =0$ is incoming from the right. The magnetic layer 
is represented by a barrier of height $V$ and width $L$; the overlayer 
has a thickness $T$, and the vacuum is modelized by a semi-inifinite 
potential barrier of height $W$. The spin dependence of the magnetic 
barrier is not considered here.

Since the energy of the incoming electron is smaller than the vacuum 
barrier height, particle flux conservation imposes to have $|r|=1$, 
and the reflection coefficient may be written as $r=e^{i\phi}$. It is 
thus sufficient to discuss the behavior of the reflection phase shift 
$\phi$. The variation of $\phi$ as a function of the overlayer thickness 
$T$ is shown in Fig.~\ref{ fig_refl}. The behavior is very contrasted, 
depending on the value of $V$. Two limit cases can be considered.

In the first case, $V$ is small as compared to $\varepsilon$, the reflection 
phase shift is essentially given by the sum of 
the phase shift due to the reflection on vacuum and and of the one due 
to the round trip through the magnetic layer and the overlayer, so 
that one has a linear variation of $\phi$ versus $T$, as appears in 
Fig.~\ref{ fig_refl} for $V=0$ and $V=0.5$~eV. Thus, the reflection 
coefficient $r=e^{i\phi}$ varies periodically versus overlayer 
thickness $T$, with a period equal to $2\pi /q_\bot$. The change of 
phase shift as $V$ varies is due to the change in the phase shift 
associated with the travel through the magnetic layer.

The opposit limit case is obtained when $\varepsilon <V$; in this case, 
$r$ is essentially constant, exepct for resonances where $\phi$ 
makes a jump of $2\pi$, corresponding to the crossing quasi-bound 
states in the overlayer. This clearly demonstrated in Fig.~\ref{ 
fig_refl}, for $V=2.0$~eV and $V=2.5$~eV. As the transmission through 
the magnetic barrier decreases, the resonances become narrower and 
their effect is completely negligible.

The intermediate situation evolves continuously between the two limit 
cases, as the curves corresponding to $V=1.0$~eV and $V=1.5$~eV in 
Fig.~\ref{ fig_refl} show.

\begin{figure}
\includegraphics{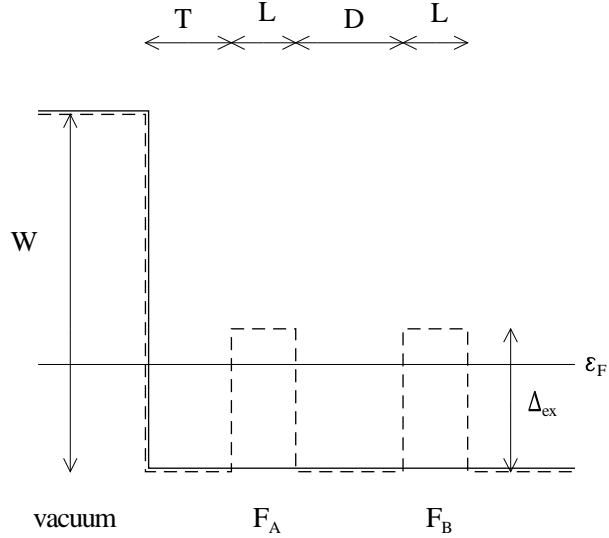}
\vspace*{8cm}
\caption{Sketch of the model used to discuss the influence of the overlayer 
and vacuum barrier on the interlayer exchange coupling; the solid and dashed 
line represent, respectively, the majority spin and minority spin potential 
profiles, for the configuration of ferromagnetic alignement.}
\label{ fig_model2}
\end{figure}

\begin{figure}[p]
\includegraphics{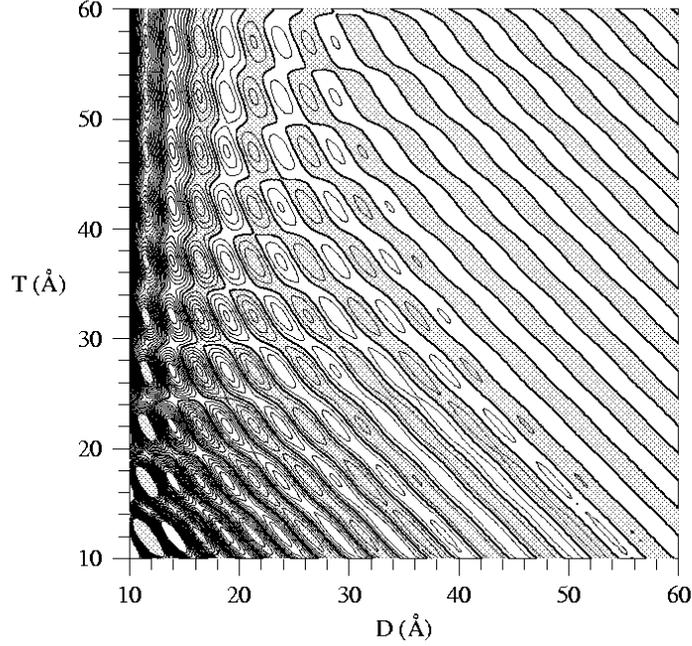}
\vspace*{9.8cm}
\includegraphics{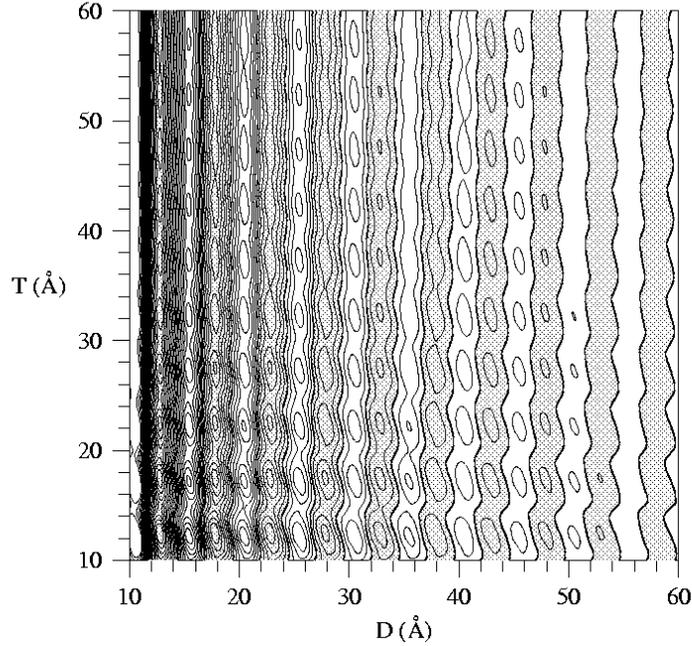}
\vspace*{11.8cm}
\caption{Contour plot of the interlayer exchange coupling, $E_F-E_{\mbox{AF}}$, 
versus spacer thickness $D$ and overlayer thickness $T$. Parameters: top pannel:
$\varepsilon_F = 1.5$~eV, $W = 5.0$~eV, 
$\Delta_{\mbox{\scriptsize ex}} = 0.5$~eV, $L=6.0$~\AA ; bottom pannel:
$\varepsilon_F = 1.5$~eV, $W = 5.0$~eV, 
$\Delta_{\mbox{\scriptsize ex}} = 2.0$~eV, $L=6.0$~\AA . The spacing between 
succesive contour lines is repectively $2.0\times 10^{-3}$~erg.cm${}^{-2}$
and 0.2~erg.cm${}^{-2}$ in the top and bottom pannels; the shaded 
area corresponds to antiferromagnetic coupling.}
\label{ fig_overl}
\end{figure}

With help of the above argument, we can understand easily how the thickness 
of the overlayer influences the IEC. As will be discussed below, the 
variation of the interlayer coupling 
versus overlayer thickness $T$ has a completely different behavior, 
depending on wether the magnetic barrier strongly confines the 
minority spin electrons at Fermi level(i.e., $\varepsilon_F < 
\Delta_{\mbox{\scriptsize ex}}$) or not; for brievety, in the 
following these two opposit situations will be refered to as strong 
confinement and weak confinement, respectively.

Let us consider first the case of weak confinement. In this case, one finds that 
the IEC dependence upon overlayer thickness $T$ is via an oscillatory function 
of $T+D$, as appears clearly from Fig.~\ref{ fig_overl} (top).

On the other hand, for the strong confinement case, as Fig.~\ref{ fig_overl} 
(bottom)  
shows, the interferences in the overlayer have only a small influence and 
yields a small modulation of the coupling strength as $T$ varies.

\begin{table}
\caption{Comparison between the theoretical predictions of 
Ref.~\protect\cite{ Bruno1996} and 
experimental observations for the oscillation periods of interlayer 
exchange coupling versus overlayer thickness.}
\label{ tab_comp}
\vspace*{0.5\baselineskip}
\begin{tabular}{ccccc}
overlayer & theoretical periods & system & experimental periods & Ref. 
\\
\tableline
Cu(001) & $\Lambda_1 =$\dec 2.6 AL & Cu/Co/Cu/Co/Cu(001) & $\Lambda 
\approx$\dec 5. AL & \cite{ Vries1995} \\
        & $\Lambda_2 =$\dec 5.9 AL &                      &      & \\
\tableline
Au(001) & $\Lambda_1 =$\dec 2.5 AL & Au/Fe/Au/Fe/Au(001) & $\Lambda_1 
\approx$\dec 2.6 AL & \cite{ Okuno1995} \\
        & $\Lambda_2 =$\dec 8.6 AL &                      & $\Lambda_2 
\approx$\dec 8.0 AL & \\
\tableline
Au(111) & $\Lambda =$\dec 4.8 AL & Au/Co/Au/Co/Au(111) & $\Lambda 
\approx $\dec 5. AL & \cite{ Bounouh1996} \\
\end{tabular}
\end{table}

This effect, which follows directly from the quantum 
interference (or quantum size effect) mechanism, has been proposed and 
experimentally confirmed independently by de Vries {\em et al.\/} 
\cite{ Vries1995} for the Co/Cu/Co(001) system with a Cu(001) overlayer, 
by Okuno and Inomata \cite{ Okuno1995} for the Fe/Au/Fe(001) system with a 
Au(001) overlayer, and by Bounouh {\em et al.\/} \cite{ Bounouh1996} for the 
Co/Au/Co(0001) with a Au(111) overlayer. In both cases, the observed period(s) 
for the oscillations versus overlayer thickness were found to be in good 
agreement with the theoretically predicted ones. This effect has also been 
confirmed by means of first-principles calculations for the Co/Cu/Co(001) with 
a Cu overlayer \cite{ Kudrnovsky1997}. The comparison between the periods of 
oscillations versus overlayer thickness predicted theoretically and those 
observed experimentally is given in Table~\ref{ tab_comp}.

\section{Exchange interactions within density functional theory}

The mechanism of IEC presented above is based upon an independent electron 
picture. That such a picture is actually valid is not immediatly obvious, in 
view of the fact that exchange interactions are indeed due to the Coulomb 
interaction between electrons. In this Section, we present the justification 
of the independent electron picture for the IEC. It relies on the use of 
``force theorems'' for the magnetic interactions, which follow from variational 
properties of various energy functionals. Since these considerations are not 
particular to the problem of interlayer coupling, but apply to any kind of 
exchange interactions in metallic magnetic systems, a more general 
point of view will be adopted here.

\subsection{Constrained density functional theory}

Almost all modern methods of electronic structure calculation rely the 
density functional theory (DFT) of Hohenberg and Kohn \cite{ Hohenberg1964}, 
within the local density approximation (LDA) of Kohn and Sham \cite{ Kohn1965}. 
In the DFT (generalized to take the spin polarization into account 
\cite{ Barth1972}) the system under consideration is described in terms of the 
local density spinor $\tn(\br)$ whose matrix elements are given 
by\footnote{Throughout this paper the tilde $\tilde{\ }$ will be used to note 
spinor quantities.}
\begin{equation}
\tn_{\alpha\beta} (\br ) = \left< \Psi \left| \psi^\dag_\alpha 
(\br ) \psi_\beta (\br ) \right| \Psi \right> .
\end{equation}
In the above equation $\psi^\dag_\alpha (\br )$ ($\psi_\beta (\br )$) is 
the second-quantization 
creation (destruction) operator for an electron of spin $\alpha$ ($\beta$) 
at $\br $, and $\left| \Psi \right>$ is the many-body wave function. 
Alternatively, the system can be described in terms of the more familiar 
local charge and spin densities, $\rho (\br )$ and ${\bf m}(\br )$, which 
are related to the density spinor by
\begin{eqnarray}
\rho &=& \mbox{tr} \left( \tn \right)  \\
{\bf m} &=& \mbox{tr} \left( \tn \, \tbsigma \right) \\
\tn  &=& \frac{\left( \rho\, \tsigma_0 + {\bf m} \cdot \tbsigma 
\right)}{2}
\end{eqnarray}
where 
\begin{equation}
\tsigma_0 \equiv \left( 
\begin{array}{cc}
1 & 0 \\
0 & 1
\end{array} \right)
\end{equation}
and 
\begin{equation}
\tbsigma \equiv \left( 
\begin{array}{cc}
\hat{{\bf z}} & \hat{{\bf x}} - {\rm i} \hat{{\bf y}} \\
\hat{{\bf x}} + {\rm i} \hat{{\bf y}}  & -\hat{{\bf z}}
\end{array} \right)
\end{equation}
is the vector whose components are the Pauli matrices; $\hat{{\bf x}}$, 
$\hat{{\bf y}}$ and $\hat{{\bf z}}$ are the unit vectors of the cartesian axes, 
and ``$\rm{tr}$'' represents the trace over spin indices.

The DFT establishes that the total energy of a given system is a unique 
functional ${\cal E}[\tn ]$ of the density spinor $\tn (\br )$ which is 
stationary and has its absolute minimum for the density spinor $\tn_0 
(\br )$ corresponding to the ground state of the system, and that the 
minimum energy is equal to the ground state energy, ${\cal E}_0$, i.e.,
\begin{equation}
{\cal E}_0 = {\cal E}[ \tn_0 ] = {\rm min}_{\,\tn }\, {\cal E}[ \tn ] .
\end{equation}
The approach universally used to compute the ground state density spinor 
$\tn_0$ and the ground state energy ${\cal E}_0$ is the one proposed 
by Kohn and 
Sham \cite{ Kohn1965}, who rewrote the energy functional as
\begin{equation}
{\cal E}[ \tn ] = {\cal T}_0 [ \tn ] + {\cal V}_{\rm ext}[ \tn ] 
+ {\cal U}_H [ \tn ] + 
{\cal E}_{\rm xc} [ \tn ] . 
\end{equation}
In the above equation, the first term is the kinetic energy of a fictitious 
system of independent electrons having the same density spinor $\tn 
(\br )$ as the real system. The second and third terms are respectively 
the potential energy of the electrons in the external potential and the 
Hartree approximation to the energy of Coulomb repulsion between electrons. 
The last term represents the exchange and correlation correction to the 
kinetic and Coulomb energies. 

The Hartree and external potential terms can be calculated trivially. The 
exchange-correlation term cannot be calculated exactly; however, convenient 
and efficient approximations exist for computing it within the LDA. The most 
problematic term is the kinetic energy ${\cal T}_0[\tn]$, for which no 
satisfactory approximation is known. Kohn and Sham solved the problem by 
showing that the system can be mapped to a fictitious system of independent 
electrons having the same density spinor and moving in an effective local 
potential spinor given by
\begin{equation}
\tw_{\rm eff} [ \tn ](\br) \equiv \tw_{\rm ext}(\br) + \tw_H [ \tn ](\br) + 
\tw_{\rm xc} [ \tn ](\br) 
\end{equation}
where $\tw_{\rm ext}(\br)$ is the external potential spinor, and
\begin{eqnarray}
\tw_H [ \tn ](\br) &\equiv& \frac{\delta {\cal U}_H[\tn ]}{\delta \tn 
(\br )} \\
\tw_{\rm xc} [\tn ](\br) &\equiv& \frac{\delta {\cal E}_{\rm xc}[\tn ]}
{\delta \tn (\br )}
\end{eqnarray}
are respectively the Hartree and exchange-correlation potential spinors. 
One has to solve Schr\"odinger-like equations for independent electrons in 
the effective potential spinor $\tw_{\rm eff}[\tn]$ (Kohn-Sham equations). 
The ground state energy is then given by
\begin{equation}
{\cal E}_0 = \sum_{\varepsilon_i \le \varepsilon_F} \varepsilon_i + 
{\cal V}_{\rm ext}[\tn_0] + {\cal U}_H[\tn_0]+ {\cal E}_{\rm xc}[\tn_0] 
- \int\! \rmd^3\br  \, \mbox{tr} \left( \tn_0\, \tw_{\rm eff}[\tn_0] \right)  
\end{equation}
where $\varepsilon_i$ are the single particle energies corresponding to the 
solutions of the Kohn-Sham equations. Since the effective potential spinor 
depends on the density spinor, the Kohn-Sham equations must be solved 
self-consistently, which is usually achieved iteratively, starting from a 
trial density spinor.

As it stands, this theory is not very convenient for computing exchange 
interactions: indeed, as it gives some information only on the ground state 
of a system, one would learn only which configuration of magnetic moments 
(e.g., ferromagnetic, antiferromagnetic, or canted) corresponds to the ground 
state; thus would obtain the {\em sign} of the interaction, but one would not 
learn anything about its {\em strength}. For this, one would have to include 
in the Hamiltonian the Zeeman interaction due to an external magnetic field, 
and compute the configuration of magnetic moments, and the total energy of the 
system as a function of this external field. This precisely how one proceeds 
in an experiment! Although this is a conceptually straightforward approach, 
it is not very convenient to implement, and has never been actually used.

Actually, what we would like to know is the ground state energy of the system, 
subjected to the restriction that the local spin-polarization ${\bf m}(\br )$ 
is constrained to be along some prescribed direction of unit vector 
$\bu (\br )$; the constraint may be extended to the whole space or 
restricted to a given subspace $V$. This approach relies on a particular 
case of the constrained density functional theory (CDFT) of Dederichs 
{\em et al} \cite{ Dederichs1984}. It uses the standard method of Lagrange 
transformation, in which one defines the new functional
\begin{equation}
{\cal F}_{\bu} [\tn ,\h ] \equiv {\cal E}[\tn ] - \int_V \! \rmd^3\br  \ 
{\bf m}(\br )\cdot \h (\br ) .
\end{equation}
The Lagrange parameter ${\bf h}_\bot$ is a transverse magnetic field, 
perpendicular to the local magnetization direction $\hat{{\bf u}}$, to be 
determined self-constently. The density spinor of the constrained ground 
state $\tn^\star_{\bu}(\br $, its energy ${\cal E}_{\rm exch}[ \bu ]$, and the 
corresponding transverse field $\h^\star$ are obtained by minimizing 
${\cal F}_{\bu}[ \tn , \h ]$ with respect to $\tn$ and $\h$, i.e.,    
\begin{eqnarray}
{\cal E}[ \tn^\star_{\bu} ] &=& {\rm min}_{\,\tn , \h}\, {\cal F}_{\bu} 
[ \tn , \h ] = 
{\cal F}_{\bu} [ \tn^\star_{\bu} , \h^\star ]\\
&\equiv& {\cal E}_{\rm exch} [ \bu ] 
\end{eqnarray}
The physical meaning of $\h (\br )$ is quite clear: it is the transverse 
external field one needs to adjust in order to maintain everywhere the 
magnetization parallel to the prescribed direction $\bu (\br )$. The local 
density of torque due to exchange interactions upon imposing the constraint 
$\bu (\br )$ is given by
\begin{equation}
\bGamma_{\bu} (\br ) \equiv -\, \frac{\delta {\cal E}_{\rm exch}
\left[ \hat{{\bf u}} 
\right] }{\delta \bOmega (\br )} 
\end{equation}
where the vector $\bOmega (\br )$ represents a local rotation of the 
spin-polarization axis. It is related to the transverse field by
\begin{equation}\label{eq:th:torque}
\bGamma_{\bu}(\br ) = -\,{\bf m}_{\bu}(\br ) \times \h .
\end{equation}

The constrained ground state density spinor $\tn^\star_{\bu}$ is 
calculated as in the unconstrained DFT, by solving self-consistently the 
Kohn-Sham equations with the effective potential spinor
\begin{equation}
\tw_{\rm eff} [ \tn ,\h ](\br) = \tw_{\rm ext}(\br) + \tw_H [ \tn ](\br) + 
\tw_{\rm xc} [ \tn ](\br) 
- \h(\br) \cdot \tbsigma .
\end{equation}
The constrained ground state energy is then calculated as
\begin{eqnarray}
{\cal E}_{\rm exch}[\bu ] &=& \sum_{\varepsilon_i \le \varepsilon_F} 
\varepsilon_i - 
\int\! \rmd^3\br  \, \mbox{tr} \left( \tn_{\bu}^\star \,
\tw_{\rm eff}[\tn_{\bu}^\star ,\h^\star ] \right) 
+ {\cal V}_{\rm ext}[\tn_{\bu}^\star ] + {\cal U}_H[\tn_{\bu}^\star ]+ 
{\cal E}_{\rm xc}[\tn_{\bu}^\star ] \nonumber \\
&&- \int_V \! \rmd^3\br  
\ {\bf m}_{\bu}^\star 
\cdot \h^\star  .
\end{eqnarray}
The functional ${\cal E}_{\rm exch}[ \bu ]$, or, equivalently, 
the torque density 
$\bGamma_{\bu} (\br )$, contains all the information we may wish to know 
about exchange interactions in the system.

\subsection{Harris-Foulkes functional and ``force theorems''}

The self-consistent method described above constitute a conceptually 
straightforward approach for computing exchange interactions. However, this 
approach also has severe drawbacks. 
\begin{itemize}
\item First of all, self-consistent calculations are computationally very 
demanding, because a large number of iterations are usually required to 
achieve convergence towards the self-consistent solution with sufficient 
accuracy. 
\item Second, within the self-consistent approach the exchange energy, which 
is typically of the order of $10^{-4}$ to $10^{-3}$ eV$/$atom, is obtained as 
the difference between the total energies for two different configurations, 
which are of the order of $10^4$ eV/atom; thus, the total energies must be 
obtained with a tremenduous relative accuracy in order to get reliable 
results. In this case, the results may be plagued by roundoff errors.
\item But the most serious difficulty we face when using the self-consistent 
approach is that it provides us very little physical insight about the 
mechanism of exchange interaction.
\end{itemize}

For all the reasons mentioned above, it is desirable to develop a method 
which is computationally convenient and accurate, and at the same provides a 
clear physical picture of the mechanism of exchange interaction. The fact the 
energy of exchange interactions is small as compared to the total energy of 
the system suggests us that a perturbation-like theory might be appropriate. 
Alternatively, one can exploit the variational properties of the density 
functional to compute approximately the constrained ground state energy in a 
single non-self-consistent shoot. This the approach we shall adopt here. We 
shall derive a force theorem \cite{ Mackintosh1980, Weinert1985} that allows 
to express ${\cal E}_{\rm exch}[\bu ]$ (within an unimportant constant) in 
terms of 
the sum of single-particle energies, calculated non-self-consistently.

Because of its variational property of the energy fonctional ${\cal F}_{\bu} 
[\tn ,\h ]$ satisfies
\begin{equation}
{\cal F}_{\bu} [\tn ,\h ] = {\cal E}_{\rm exch}[\bu ] + {\cal O}_2 \left( \tn - 
\tn_{\bu}^\star , \h - \h^\star \right)
\end{equation}
where ${\cal O}_2 \left( \tn - \tn_{\bu}^\star , \h - \h^\star \right)$ is 
of second order with respect to $\tn - \tn_{\bu}^\star$ and $\h - 
\h^\star$. So, if we have good guesses $\tn$ and $\h$ for the exact 
$\tn_{\bu}^\star$ and $\h^\star$, and if we are able to compute 
${\cal F}_{\bu} [\tn ,\h ]$, we get an estimate of ${\cal E}_{\rm exch}[\bu ]$ 
which is accurate 
to second order in the error of our initial guesses. As we shall see just 
below, the difficulty of this approach lies in the calculation of 
${\cal F}_{\bu}[\tn ,\h ]$ for the chosen $\tn$ and $\h$.

This is seen as follows. From a trial input density $\tn_{\rm in}$ and 
input transverse field $\h^{\rm in}$ one gets the effective potential spinor 
$\tw_{\rm eff}[ \tn_{\rm in}, \h^{\rm in}]$; by solving the Kohn-Sham 
equations for this effective potential spinor one gets a set of single 
particle energies $\varepsilon_i [\tn_{\rm in}, \h^{\rm in}]$, from which 
one can in turn compute an output density $\tn_{\rm out}$. From this, we 
can compute
\begin{eqnarray}
{\cal F}_{\bu}[\tn_{\rm out},\h^{\rm in} ] &=& \sum_{\varepsilon_i \le 
\varepsilon_F} \varepsilon_i [\tn_{\rm in}, \h^{\rm in}] 
- \int\! \rmd^3\br \, \mbox{tr} \left( \tn_{\rm out} \,
\tw_{\rm eff}[\tn_{\rm in},\h^{\rm in}] \right) \nonumber \\
&&+ {\cal V}_{\rm ext}[\tn_{\rm out}] + {\cal U}_H[\tn_{\rm out}]+ 
{\cal E}_{\rm xc}[\tn_{\rm out}] 
- \int_V \! \rmd^3\br \ 
{\bf m}_{\rm out}\cdot \h^{\rm in}  
\end{eqnarray}
We now see where the problem lies: we get an estimate of the energy 
fonctional corresponding to the output density spinor, not to the input one; 
furthermore, the expression mixes in a complicated manner $\tn_{\rm in}$ 
and $\tn_{\rm out}$. Since we don't know the input density spinor 
corresponding to an arbitrary output density spinor, this cannot be used 
to compute the energy fonctional for a chosen density spinor.

This problem was circumvented by Harris \cite{ Harris1985} who defined an 
auxiliary functional as follows
\begin{eqnarray}\label{eq:th:Harris}
{\cal{G}}_{\bu}[\tn_{\rm in}, \tw_{\rm in} , \h^{\rm in}] &\equiv& 
\sum_{\varepsilon_i \le \varepsilon_F} \varepsilon_i[\tw_{\rm in}] 
- \int\! \rmd^3\br \, \mbox{tr} \left( \tn_{\rm in}\, \tw_{\rm in} \right) 
+ {\cal V}_{\rm ext}[\tn_{\rm in}] + {\cal U}_H[\tn_{\rm in}]+ 
{\cal E}_{\rm xc}[\tn_{\rm in}] \nonumber \\
&&- \int_V \! \rmd^3\br\ {\bf m}_{\rm in} 
\cdot \h^{\rm in}
\end{eqnarray}
where $\tn_{\rm in}$ is the input density, $\varepsilon_i [\tw_{\rm in}]$ 
are the single particle energies calculated with the input effective potenti 
$\tw_{\rm in}$, and $\h^{\rm in}$ is the input transverse field. Here we 
have used a generalization of the Harris functional proposed by Foulkes and 
Haydock \cite{ Foulkes1989}. It is straighforward to show that
\begin{equation}
{\cal{G}}_{\bu} [\tn , \tw ,\h ] = {\cal F}_{\bu}[\tn^\star_{\bu}, \h^\star ] 
+ {\cal O}_2 \left( \tn - \tn_{\bu}^\star , \tw - \tw_{\bu}^\star , \h 
- \h^\star \right)
\end{equation}
where 
\begin{equation}
\tw_{\bu}^\star \equiv \tw_{\rm ext} + \tw_H [\tn^\star_{\bu}] + 
\tw_{\rm xc} [\tn^\star_{\bu}] - \h^\star \cdot \tbsigma 
\end{equation}
and ${\cal O}_2 \left( \tn - \tn_{\bu}^\star , \tw - \tw_{\bu}^\star , \h 
- \h^\star \right)$ is a (non generally positive) error of second order with 
respect $\tn - \tn_{\bu}^\star$, $\tw - \tw_{\bu}^\star$ and $\h 
- \h^\star$. The properties of the Harris functionals have been discussed by 
a number of authors \cite{ Read1989, Finnis1990, Robertson1991} who found that 
it often yields better approximate estimations of the ground state energy than 
the Hohenberg-Kohn functional.

We now have all the material needed to establish the force theorem for magnetic 
exchange interactions. To this end, we make the following choice for the input 
values of $\tn$, $\tw$ and $\h$:
\begin{mathletters}
\begin{eqnarray}\label{eq:th:input}
\tn_{\rm in}(\br) &=& \frac{\rho_{\rm in}(\br)\, \tsigma_0 
+ m_{\rm in}(\br)\, \bu(\br) \cdot 
\tbsigma}{2} \\
\h^{\rm in}(\br) &=& 0 \\
\tw_{\rm in}(\br) &=& v_{\rm in}(\br) \,\tsigma_0 - h_{\rm in}(\br)\, 
\bu (\br)\cdot \tbsigma .
\end{eqnarray}
\end{mathletters}
The input charge and spin densities are chosen to be independent of $\bu$ 
in magnitude, with the axis of the spin-polarization along $\bu$; thus we 
neglect the redistribution of charge and magnetic moment due to rotating the 
magnetic moments, which usually constitutes a good approximation. The input 
effective field is chosen to be parallel to $\bu$; this is usually a good 
approximation, because the transverse component of the effective field is 
much smaller than its longitudinal component, which is of the order of the 
exchange splitting. 

For the magnitude of the input effective potential $v_{\rm in}$ and input 
effective field $h_{\rm in}$ we can specify a little more by ascribing them 
the value corresponding to the LDA, i.e.,
\begin{mathletters}
\begin{eqnarray}
v_{\rm in}(\br) &=& v_{\rm ext}(\br) + v_H[\rho_{\rm in}](\br) + 
v_{\rm xc}^{\rm LDA}[\rho_{\rm in}, m_{\rm in}](\br) \\
h_{\rm in}(\br) &=& h_{\rm xc}^{\rm LDA} [\rho_{\rm in}, m_{\rm in}](\br) .
\end{eqnarray}
\end{mathletters}
However, this choice is not necessary and the stationarity of the Harris 
functional allows more flexibility in the choice of the input values for 
the effective potential and effective field. 

Inserting the particular choice (\ref{eq:th:input},\textit{b,c}) for the 
input quantities in the definition of the Harris functional 
(\ref{eq:th:Harris}) we find
\begin{eqnarray}
{\cal{G}}_{\bu}[\tn_{\rm in}, \tw_{\rm in} , \h^{\rm in}] &\equiv&  
\sum_{\varepsilon_i \le \varepsilon_F} \varepsilon_i 
[v_{\rm in}, h_{\rm in}\bu ]
- \int\! \rmd^3\br \,  
\left( \rho_{\rm in} v_{\rm in} - m_{\rm in} h_{\rm in} \right) \nonumber \\
&&+ {\cal V}_{\rm ext}[\rho_{\rm in}] + {\cal U}_H[\rho_{\rm in}]+ 
{\cal E}_{\rm xc}[\rho_{\rm in} , m_{\rm in}] .
\end{eqnarray}
We find that the only term which depends on the constraint $\bu$ is the sum 
of single particle energies. Thus, we finally obtain
\begin{eqnarray}
{\cal E}_{\rm exch}[{\bu}_1] - {\cal E}_{\rm exch}[{\bu}_2] &=& 
\sum_{\varepsilon_i \le 
\varepsilon_F} \varepsilon_i [v_{\rm in}, h_{\rm in}\bu_1 ] 
- \sum_{\varepsilon_i \le \varepsilon_F} \varepsilon_i 
[v_{\rm in}, h_{\rm in}\bu_2 ] \nonumber \\
&&+ {\cal O}_2 \left( \rho_{\rm in} - \rho_{\bu}^\star, m_{\rm in} 
- m_{\bu}^\star, \h^\star \right)
\end{eqnarray}
which consitutes the force theorem for the exchange interaction energy. 
This is an important result, for several reason:
\begin{itemize}
\item The energy of exchange interactions ${\cal E}_{\rm exch}[\bu ]$ can be 
computed (within an unimportant constant) in a single shoot, without need 
for lengthy iterations towards self-consistently, which makes the calculations 
considerably faster and easier.
\item Although the exchange energy is ultimately due to Coulomb interactions 
between electrons, it is expressed here as a sum of single-particle energies 
for indenpendent electrons. This is of considerable practical importance, 
because the sum of single particle energies is much smaller than the total 
energy of the system; thus the relative accuracy needed is much less than 
for self-consistent total energy calculations, and the risk of computational 
error is much smaller.
\item A further remarkable feature of this result is that the exchange 
energy difference depends only on the input potential and input field, and 
not at all on the input charge and spin densities (as long as the latter 
don't vary much upon rotating the moments). This leaves considerable 
flexibility in setting up practical computational schemes, and allows to 
use suitably parametrized schemes, without jeopardizing seriously the 
accuracy of the results.
\end{itemize}

The torque density is given by 
\begin{eqnarray}
\bGamma_{\bu} (\br) &\equiv& -\, \frac{\delta {\cal E}_{\rm exch}[\bu ]}
{\delta \bOmega (\br)} \nonumber \\
&\approx& -\, \sum_{\varepsilon_i \le \varepsilon_F} 
\frac{\delta \varepsilon_i [\bu ]}{\delta \bOmega (\br)} .
\end{eqnarray}
From the Hellman-Feynman theorem, we obtain for a given single-particle 
energy $\varepsilon_i [\bu ]$ corresponding to the wave function 
$\left| \phi_i [\bu ]\right>$,
\begin{equation}
\frac{\delta \varepsilon_i [\bu ]}{\delta \bOmega (\br)} 
= \left< \phi_i [\bu ] \right| \frac{\delta H}{\delta\bOmega (\br)} 
\left| \phi_i [\bu ] \right> .
\end{equation}
Upon performing the rotation $\bOmega (\br) \equiv \Omega(\br) \,
{\bf n}_\bOmega(\br)$, 
which changes 
$\bu (\br)$ into $\bu^\prime (\br)$, the effective potential 
$\tw_{\rm eff}(\br)$ 
is transformed according to (for simplicity, the subscript ``eff'' will 
be ommited below)
\begin{equation}
\tw_{\bu}(\br) \ \to \ \tw_{\bu^\prime}(\br) = 
\tR(\br)\ 
\tw_{\bu}(\br) \ \tR^{-1} (\br) 
\end{equation}
where the rotation matrix is given by
\begin{eqnarray}
\tR(\br) &\equiv& \rme^{-\rmi \bOmega (\br) \cdot \tbsigma} \nonumber \\
&=& \cos\left(\frac{\Omega (\br)}{2}\right) - \rmi\, {\bf n}_\bOmega (\br) \,
\Omega (\br) \cdot \tbsigma \, \sin\left(\frac{\Omega 
(\br)}{2}\right) .
\end{eqnarray}
Thus,
\begin{equation}
\frac{\delta  H  }{\delta\bOmega (\br)} = -\ \left| \br \right> 
h_{\rm in}(\br)\,\bu \cdot \tbsigma \ \left< \br \right|
\end{equation}
and by using the relation
\begin{equation}
{\bf m}_{\rm out}(\br) =  \sum_{\varepsilon_i \le \varepsilon_F}  
\left< \phi_i [\bu ] \, | \br \right>  \tbsigma\, 
\left< \br | \phi_i [\bu ] \right> 
\end{equation}
we finally obtain
\begin{equation}
\bGamma_{\bu} (\br) =-\ h_{\rm in}(\br)\ \bu \times  
{\bf m}_{\rm out} (\br) .
\end{equation}
The above result constitutes the force theorem for the exchange torque density.

\subsection{Examples of first-principles calculations of IEC}

The use of the magnetic ``force theorems'' discussed in the preceding Section 
allows to reduce by several orders of magnitude the computation times required 
to calculate the interlayer exchange coupling. Such first-principles 
calculations have been performed by a number of authors 
\cite{ Schilfgaarde1993, Lang1993, Nordstrom1994, Kudrnovsky1994, 
Lee1995, Kudrnovsky1996, Lang1996, Drchal1996, Stiles1996, Albuquerque1996, 
Lee1996, Mathon1997, Costa1997}.

\begin{figure}[p]
\includegraphics{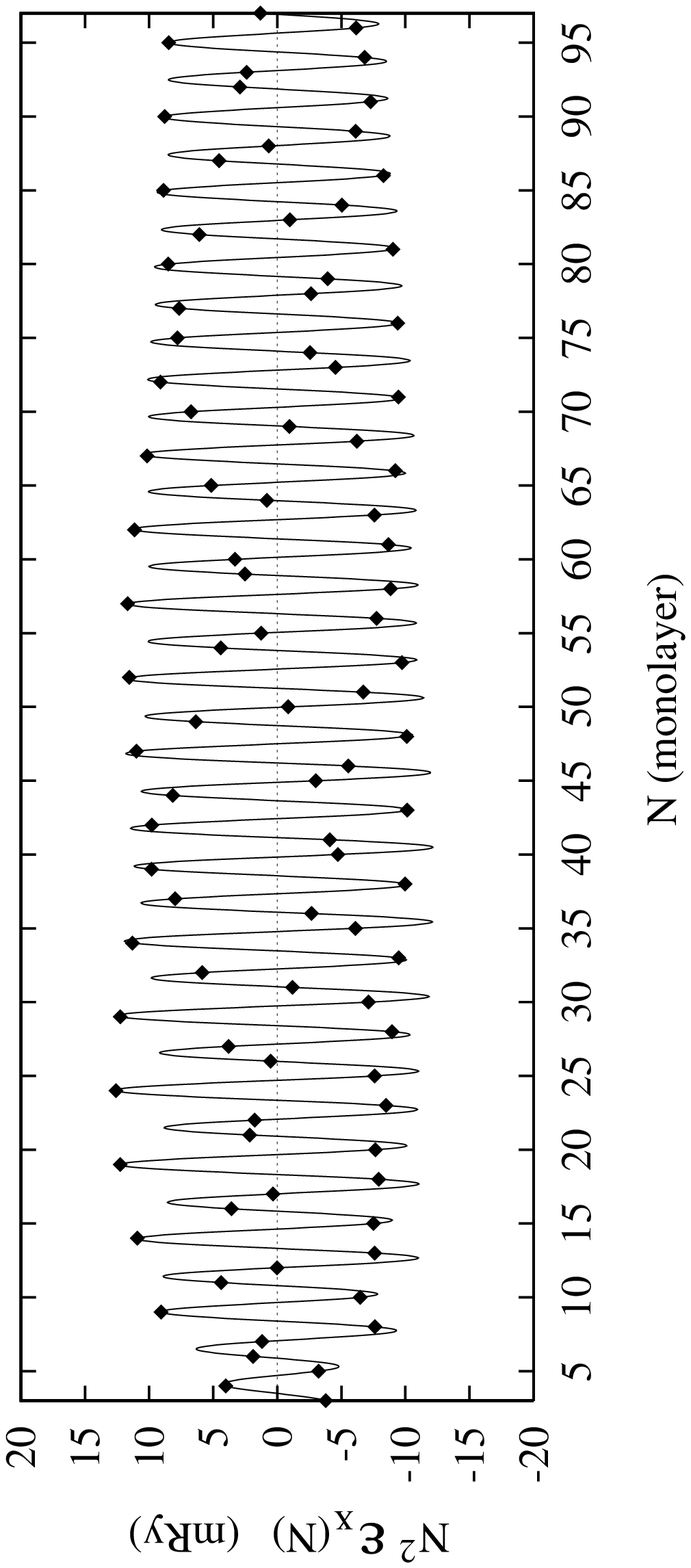}
\vspace{6.5cm}
\includegraphics{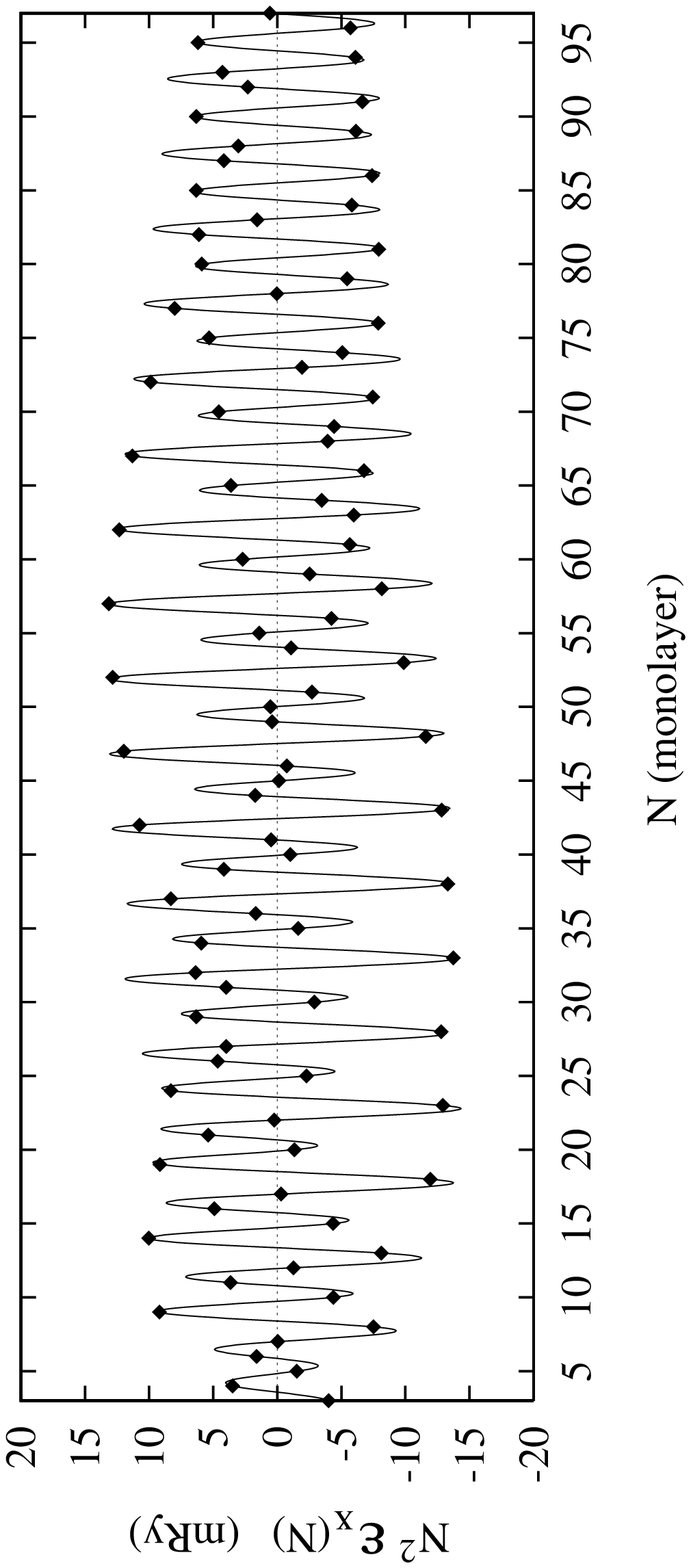}\vspace*{8cm}
\caption{Calculated interlayer exchange coupling $N^2 {\cal E}_x$ as a 
function of spacer thickness $N$ (in AL) for the Co/Cu/Co(001) system;
top panel: semi-infinite Co layers; bottom panel: 5~AL thick Co layers 
\protect\cite{ Kudrnovsky_PC}.}
\label{ fig_Co_dir}
\end{figure}

\begin{figure}[p]
\includegraphics{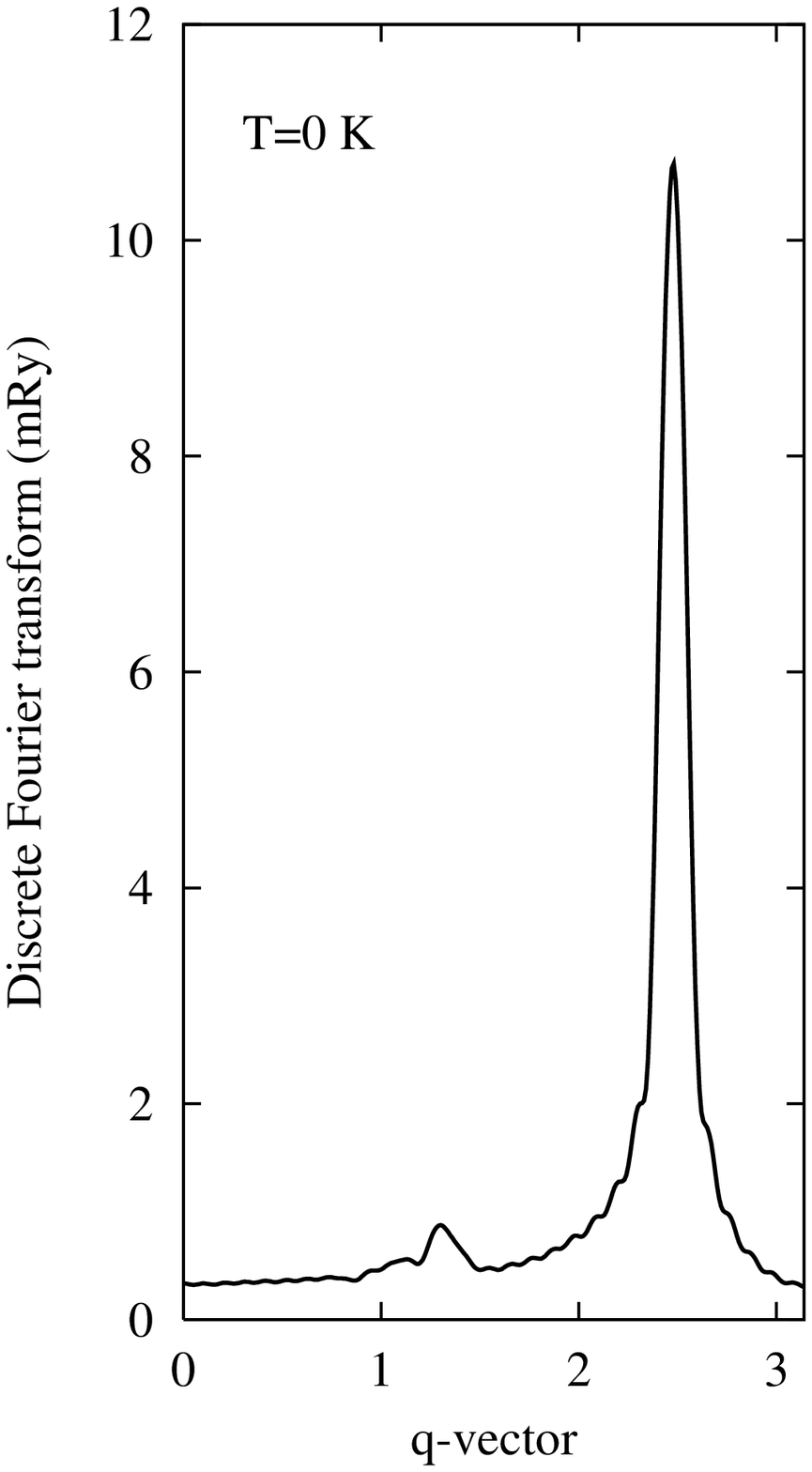}
\includegraphics{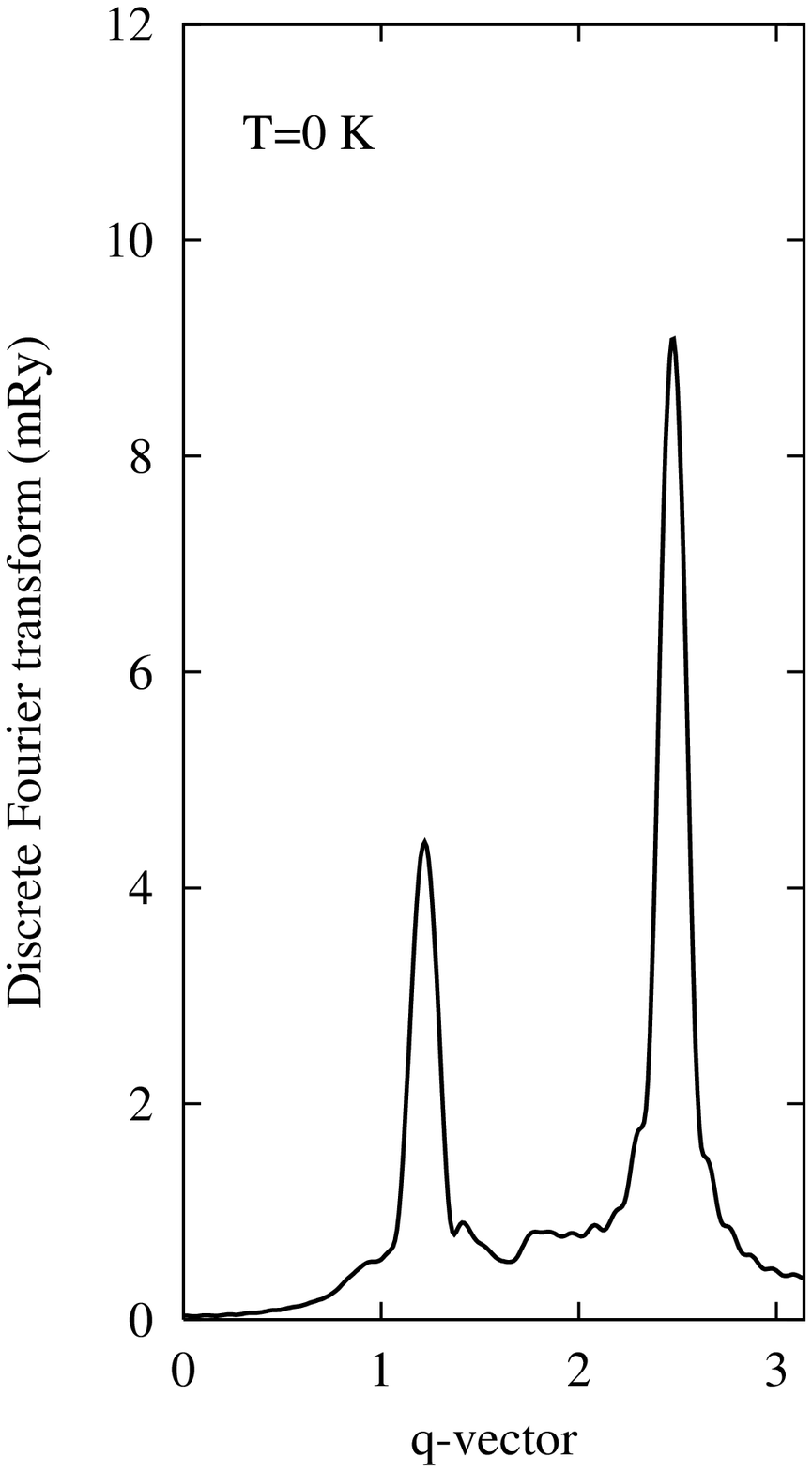}
\vspace*{12.5cm}
\caption{Discrete Fourier transform of the data shown in 
Fig.~\protect\ref{ fig_Co_dir}; left panel: semi-infinite Co layers; right panel:
5~AL thick Co layers \protect\cite{ Kudrnovsky_PC}.}
\label{ fig_Co_Fourier}
\end{figure}

The Fig.~\ref{ fig_Co_dir} shows the interlayer exchange coupling for the 
Co/Cu/Co(001) systems, for semi-infinite Co layers (top), and 5~AL thick 
Co layers (bottom) \cite{ Kudrnovsky_PC}. The data have been multiplied by 
$N^2$ in order to highlight the $1/N^2$ dependence of the IEC, as predicted from 
the asymptotic approximation. The data corresponding to the two different Co 
thicknesses seem very different at first sight. In order to evidence more 
clearly the periodic oscillatory behavior, one can perform a discrete 
Fourier analysis of the $N$ dpendence of the IEC \cite{ Kudrnovsky1994}. The 
corresponding results are shown in Fig.~\ref{ fig_Co_Fourier}. 

One sees very clearly that the oscillations comprise two superimposed 
oscillatory components, as expected, with periods in good agreement with the 
ones predicted in terms of the spacer Fermi surface \cite{ Bruno1991}. On the 
other hand the amplitude of the oscillations change dramatically, as the Co 
thickness is varied.

These results 
provide a very clear confirmation of the results obtained from the discussion
in terms of quantum interferences in the asymptotic regime, namely that the 
periods of oscillation versus spacer thickness depend only on the spacer 
material, but the amplitudes can vary with the thickness (as in the present 
case) or nature of the ferromagnetic layers.

\section{Substitutional disorder}

The case where (some part of) the system under consideration consists of 
a substitutionally disordered alloys is extremely important in practice. This 
happens of course if some layers are {\em by purpose\/} chosen to consist of 
alloy; but more generally, unavoidable interdiffusion takes place at the 
interfaces, giving rise to a disordered interface region with a progressive 
variation of concentration.

\subsection{``Vertex cancellation'' theorem}

We present here a general discussion of exchange interactions in the presence 
of substitutional disorder. The results given here are used in the present 
paper to study interlayer exchange interactions, but they are also applicable 
for studying exchange interactions {\em within\/} a ferromagnet, exchange 
stiffnesses, spin-wave energies, etc. 

The principal result is the ``vertex cancellation theorem'' of Bruno {\em et 
al.\/} \cite{ Bruno1996b}. We present here an alternative, more general, 
derivation of this result.

Let us specify the notations used here. Our purpose is to compute the total 
energy (more precisely the thermodynamic grand-potential) as a function of 
the prescribed local direction $\bu(\br)$ of the magnetization. Explicitely, 
the Hamiltonian operator corresponding to a particular configuration 
$\bu \equiv \{\bu(\br)\}$ of the local moments is written as
\begin{equation}
{\sf H} = {\sf K} + {\sf V}_\bu .
\end{equation}
The matrix elements of the kinetic energy operator ${\sf K}$ are
\begin{equation}
\left< \br \right| {\sf K} \left| \br^\prime \right> \equiv 
\delta(\br - \br^\prime)\, 
\frac{-\hbar^2}{2m}\, 
\frac{{\rm d}^2}{{\rm d}\br^2}\, \,
\tilde{\sigma}_0 ,
\end{equation}
and those of the potential operator $\sf V$
\begin{equation}
\left< \br \left| {\sf V}_\bu \right| \br^\prime \right> 
\equiv \delta(\br - \br^\prime) \, \left[ v(\br) \, \tilde{\sigma}_0 
- h(\br)\, \bu (\br) \cdot \tilde{\bsigma} \right] ,
\end{equation}
where $v(\br)$ and $h(\br)$ are respectively the local effective potential
and the local effective field. The corresponding Green's function operator is 
\begin{equation}
{\sf G}_\bu (z) \equiv \left( z - {\sf H}_\bu \right)^{-1} .
\end{equation}

Because of the substitutional disorder, $v(\br)$ and $h(\br)$ have some spacial
randomness. However, within the domain of applicability of the ``force
theorem'', 
we assume that they are unchanged upon changing the moment configuration $\bu$. 

The central quantity in the theory of disordered alloys is the 
{\em configuration averaged\/} Green's function 
\begin{equation}
\overline{\sf G}_\bu(z) \equiv \left< {\sf G}_\bu(z) \right>_c 
\equiv \left[ z -{\sf K} -{\sf \Sigma}_\bu(z) \right]^{-1}
\end{equation}
where $\left< \, \dots \, \right>_c$ indicates an average over all possible 
alloy configurations, and where ${\sf\Sigma}_\bu(z)$ is the self-energy. The 
Green's function is given by
\begin{equation}
{\sf G}_\bu(z) = \overline{\sf G}_\bu(z) + \overline{\sf G}_\bu(z) 
{\sf T}_\bu(z) 
\overline{\sf G}_\bu(z) ,
\end{equation}
where the t-matrix ${\sf T}_\bu(z)$ is given by
\begin{equation}
{\sf T}_\bu(z) \equiv \left( {\sf V}_\bu - {\sf \Sigma}_\bu(z)\right) 
\left[ 1 - \overline{\sf G}_\bu(z) 
\left( {\sf V}_\bu -{\sf\Sigma}_\bu(z)\right)\right]^{-1} .
\end{equation}
>From the definition of the configuration averaged green's function, we get 
the self-consistency condition
\begin{equation}\label{eq_sc0}
\left< {\sf T}_\bu(z) \right>_c = 0 ,
\end{equation}
or equivalently,
\begin{equation}\label{eq_sc1}
\left< \left[ 1 - \overline{\sf G}_\bu(z) 
\left( {\sf V}_\bu -{\sf\Sigma}_\bu(z)\right)\right]^{-1} \right>_c = 1 .
\end{equation}
In practice, the self-energy satisfying the conditions (\ref{eq_sc0}, 
\ref{eq_sc1}) cannot be calculated exactly and one has to resort to 
approximations. The most popular approach is the CPA, in which conditions 
(\ref{eq_sc0}, \ref{eq_sc1}) are replaced by {\em on site\/} conditions, 
for all atomic sites ${\bf R}$.

If one uses the ``force theorem'', the thermodynamic grand-potential includes 
only the single-particule energies (Kohn-Sham eigenvalues) and is given by 
\begin{equation}
\Phi_\bu = -  \int_{-\infty}^{+\infty} {\rm d}\varepsilon \, f(\varepsilon) 
\, N_\bu (\varepsilon) ,
\end{equation}
where $f(\varepsilon)$ is the Fermi-Dirac function, and $N_\bu(\varepsilon)$ 
the integrated 
density of states (averaged over all possible alloy configurations) for the
local moment configuration $\bu$. The integrated density of states
is given by
\begin{equation}
N_\bu (\varepsilon)= \frac{1}{\pi} \mbox{Im Tr} \left[ \left< \ln {\sf G}(z) 
\right>_c \right]_{z=-\infty +{\rm i}0^+}^{z=\varepsilon +{\rm i}0^+} .
\end{equation}

Let us consider the quantity 
\begin{equation}
A_{\bu}(z) \equiv \mbox{Tr} \left< \ln {\sf G}_{\bu}(z) 
\right>_c .
\end{equation}
By some simple algebra, one can show that 
\begin{equation}
A_{\bu}(z)
= \mbox{Tr} \left[ \ln \overline{\sf G}_{\bu}(z) \right]
- X_{\bu}(z)
\end{equation}
where
\begin{equation}
X_{\bu}(z) \equiv \mbox{Tr}\, \left< \ln \left[ 1 - 
{\sf G}_{\bu}(z) \left( {\sf V}_{\bu} - {\sf\Sigma}_{\bu}
(z) \right) \right] \right>_c
\end{equation}
is called the {\em vertex correction\/}. This term is usually non negligible 
and is difficult to calculate. We shall show however that its variation 
upon varying $\bu$ takes a simple form.

Let us precise the variation of the Hamiltonian under varying $\bu$. If we 
chose a reference configuration $\bu_0$ for the local moment configuration, 
the configuration $\bu$ is obtained by performing locally a rotation of 
vector
\begin{equation}
\bOmega(\br) \equiv \Omega(\br) \, \hat{\bf n}_\bOmega(\br) .
\end{equation}
Explicitly, one has
\begin{equation}
\bu  \equiv (\hat{\bf n}_\bOmega\cdot\bu_0)
\, \hat{\bf n}_\bOmega + 
\cos\Omega \, (\hat{\bf n}_\bOmega \times \bu)\times \hat{\bf n}_\bOmega 
+ \sin\Omega \, (\hat{\bf n}_\bOmega\times\bu) .
\end{equation}
The potential operator ${\sf V}_\bu$ corresponding to $\bu$ is obtained 
from ${\sf V}_{\bu_0}$ as
\begin{equation}
{\sf V}_{\bu} \equiv 
{\sf R}_\bOmega \, {\sf V}_{\bu_0} \, {\sf R}^{-1}_\bOmega ,
\end{equation}
and the matrix elements of the rotation operator ${\sf R}_\bOmega$ are 
given by
\begin{eqnarray}
\left< \br \left| {\sf R}_\bOmega \right| \br^\prime \right> 
&\equiv & \delta (\br-\br^\prime)\, 
\exp \left( -\frac{\rm i}{2} \bOmega(\br) \cdot
\tilde{\bsigma} \right) \nonumber \\
&=& \delta (\br-\br^\prime)\, \left[
\cos \left(\frac{\Omega (\br)}{2}\right)\, \tilde{\sigma}_0 
-{\rm i} \sin\left( \frac{\Omega(\br)}{2}\right) \, \hat{\bf n}_{\bOmega}(\br) 
\cdot \tilde{\bsigma} \right] .
\end{eqnarray}
We shall also make of the relation
\begin{equation}
\frac{\delta\, \left({\sf R}_\bOmega {\sf B} {\sf R}^{-1}_\bOmega \right) }
{\delta\bOmega (\br)} = -\,\frac{\rm i}{2} \left[ \left|\br \right> 
\tilde{\bsigma} 
\left<\br \right| \, ; \, {\sf R}_\bOmega {\sf B} 
{\sf R}^{-1}_\bOmega \right]_- ,
\end{equation}
for an arbitrary operator ${\sf B}$. 

The derivative of $A_\bu$ (from now on, we shall omit the dependence on 
complex energy $z$) with respect to $\bOmega$ is given by
\begin{equation}
\frac{\delta A_{\bu}}{\delta \bOmega (\br)} = 
\mbox{Tr} \left[ \overline{\sf G}_{\bu} \frac{\delta{\sf\Sigma}_{\bu}}
{\delta\bOmega(\br)} \right] - \frac{\delta X_{\bu}}
{\delta\bOmega(\br)} ,
\end{equation}
with
\begin{eqnarray}
\frac{\delta X_{\bu}}{\delta\bOmega(\br)} &=& - \mbox{Tr} 
\left[ \frac{\delta \overline{\sf G}_{\bu}}{\delta\bOmega(\br)}
\left< \left( {\sf V}_{\bu} - {\sf \Sigma}_{\bu}\right) 
\left[ 1 - \overline{\sf G}_{\bu} 
\left( {\sf V}_{\bu}-{\sf\Sigma}_{\bu}\right)\right]^{-1}
\right>_c \right] \nonumber \\
&&-\mbox{Tr} \left< \overline{\sf G}_{\bu} \left( 
\frac{\delta {\sf V}_{\bu}}{\delta\bOmega(\br)} 
- \frac{\delta {\sf\Sigma}_{\bu}}{\delta\bOmega(\br)} \right) 
\left[ 1 - \overline{\sf G}_{\bu}
\left( {\sf V}_{\bu}-{\sf\Sigma}_{\bu}\right)\right]^{-1}
\right>_c .
\end{eqnarray}
The first term on the right-hand side of the above equation is zero, 
because of condition (\ref{eq_sc0}). Next, we split the self-energy in 
two parts as 
\begin{equation}
{\sf\Sigma}_{\bu} \equiv {\sf\Sigma}_{\bu}^{(1)} + 
{\sf\Sigma}_{\bu}^{(2)} ,
\end{equation}
where ${\sf\Sigma}_\bu^{(1)}$ varies with $\bOmega$ like ${\sf V}_\bu$, i.e., 
\begin{equation}
{\sf\Sigma}_{\bu}^{(1)} \equiv {\sf R}_\bOmega {\sf\Sigma}_{\bu_0} 
{\sf R}^{-1}_\bOmega .
\end{equation}
Then, we obtain
\begin{eqnarray}
\frac{\delta X_{\bu}}{\delta\bOmega(\br)} &=& 
\mbox{Tr} \left[ \overline{\sf G}_{\bu} 
\frac{\delta {\sf\Sigma}^{(2)}_{\bu}}{\delta\bOmega(\br)} 
\left< \left[ 1 - \overline{\sf G}_{\bu}
\left( {\sf V}_{\bu}-{\sf\Sigma}_{\bu}\right)\right]^{-1}
\right>_c \right] \nonumber \\
&& - \mbox{Tr} \left< \overline{\sf G}_{\bu} \, \frac{(-{\rm i})}{2} 
\left[ \left|\br\right> \bsigma \left<\br\right|\, ; \, 
\left( {\sf V}_{\bu} 
-{\sf\Sigma}^{(1)}_{\bu} \right) \right]_- 
\left[ 1 - \overline{\sf G}_{\bu}
\left( {\sf V}_{\bu}-{\sf\Sigma}_{\bu}\right)\right]^{-1} 
\right>_c 
\end{eqnarray}
By using the permutation invariance of the trace and the conditions 
(\ref{eq_sc0}, \ref{eq_sc1}), we obtain finally
\begin{equation}
\frac{\delta X_{\bu}}{\delta\bOmega(\br)} =
\mbox{Tr} \left[ \overline{\sf G}_{\bu} 
\frac{\delta {\sf\Sigma}^{(2)}_{\bu}}{\delta\bOmega(\br)} \right]
\end{equation}
and hence
\begin{equation}
\frac{\delta A_{\bu}}{\delta \bOmega (\br)} = 
\mbox{Tr} \left[ \overline{\sf G}_{\bu} 
\frac{\delta{\sf\Sigma}^{(1)}_{\bu}}
{\delta\bOmega(\br)} \right] .
\end{equation}

Thus, the torque density due to the exchange interactions is given by
\begin{eqnarray}\label{eq_torque}
{\bf\Gamma}_{\bu}(\br) &\equiv & 
-\, \frac{\delta \Phi_{\bu}}{\delta\bOmega(\br)} \nonumber \\
&=& \frac{1}{\pi} \int_{-\infty}^{+\infty} f(\varepsilon) \mbox{ Im Tr}
\left[ \overline{\sf G}_{\bu}(\varepsilon +{\rm i}0^+) 
\frac{\delta{\sf\Sigma}^{(1)}_{\bu}(\varepsilon +{\rm i}0^+)}
{\delta\bOmega(\br)} 
\right] \, {\rm d}\varepsilon .
\end{eqnarray}
This exact result constitutes the ``vertex cancellation theorem'' for the 
torque density. Its usefulness arises from the fact that the ``vertex 
corrections'' have been eliminated. A further important feature is that the 
exact torque is given in terms of the angular derivative of 
${\sf\Sigma}_\bu^{(1)}$ only, which is known explicitly because it follows
the angular variation of ${\sf V}_\bu$; the angular derivative of the 
remaining part of the self-energy, ${\sf\Sigma}_\bu^{(2)}$, which cannot be
calculated easily, has been eliminated.

In order to compute the difference of thermodynamic grand-potential between 
two local moment configurations $\bu_1$ and $\bu_2$, we use a theorem due to
Ducastelle \cite{ Ducastelle1975}, which states that the 
thermodynamic grand-potential,
considered as a functional $\tilde{\Phi}[\overline{\sf G},{\sf V}]$ of the 
{\em independent\/} variables $\overline{\sf G}$ and ${\sf V}$, is stationary 
with respect to $\overline{\sf G}$ when the latter satisfies the 
self-consistency condition (\ref{eq_sc0}, \ref{eq_sc1}). This means that a 
first-order error on $\overline{\sf G}_\bu$ gives only a second-order 
error on $\Phi_\bu$. Writing $\overline{\sf G}_\bu$ as
\begin{equation}
\overline{\sf G}_{\bu}(z) \equiv \overline{\sf G}^{(1)}_{\bu}(z)
+ \overline{\sf G}^{(2)}_{\bu}(z) 
\end{equation}
with
\begin{equation}
\overline{\sf G}^{(1)}_{\bu}(z) \equiv 
\left( z - {\sf K} - {\sf\Sigma}^{(1)}_{\bu}(z) \right)^{-1} ,
\end{equation}
we take $\overline{\sf G}^{(1)}_{\bu}$ as a trial value for computing 
$\Phi_\bu$. This can be expected to be a good approximation, provided the
condition 
\begin{equation}\label{eq_cond}
m(\br)\, \left| \frac{{\rm d}\bOmega}{{\rm d}\br} \right| \ll k_F \, \rho(\br) 
\end{equation}
(where $\rho(\br)$ and $m(\br)$ are respectively the electron and spin 
densitites) is satisfied. Replacing $\overline{\sf G}_\bu$ by 
$\overline{\sf G}_\bu^{(1)}$ in Eq.~(\ref{eq_torque}), we can now integrate 
over angles, and we get
\begin{equation}\label{eq_vtx2}
{\Phi}_{\bu_1} - {\Phi}_{\bu_2} \approx -\, \frac{1}{\pi} 
\int_{-\infty}^{+\infty} f(\varepsilon ) \mbox{ Im Tr} \left[ 
\ln\overline{\sf G}^{(1)}_{\bu_1}
(\varepsilon +{\rm i}0^+) - \ln\overline{\sf G}^{(1)}_{\bu_2} 
(\varepsilon +{\rm i}0^+) \right] ,
\end{equation}
which constitutes the ``vertex cancellation theorem'' for exchange energies.
In the derivation of the ``vertex cancellation theorem'', we have made 
use of the exact self-consistency condition (\ref{eq_sc0}, \ref{eq_sc1}); one 
can show also that the same result holds if one uses the approximate CPA 
self-consistency condition.

In the case of interlayer coupling, the condition (\ref{eq_cond}) is
satisfied even for large rotation angles, 
because ${{\rm d}\bOmega}/{{\rm d}\br}$ differs from zero only 
in a region where $m(\bf r)$ is negligible. This was confirmed by explicit 
numerical calculations in Ref.~\cite{ Bruno1996b}.

\begin{figure}[p]
\includegraphics{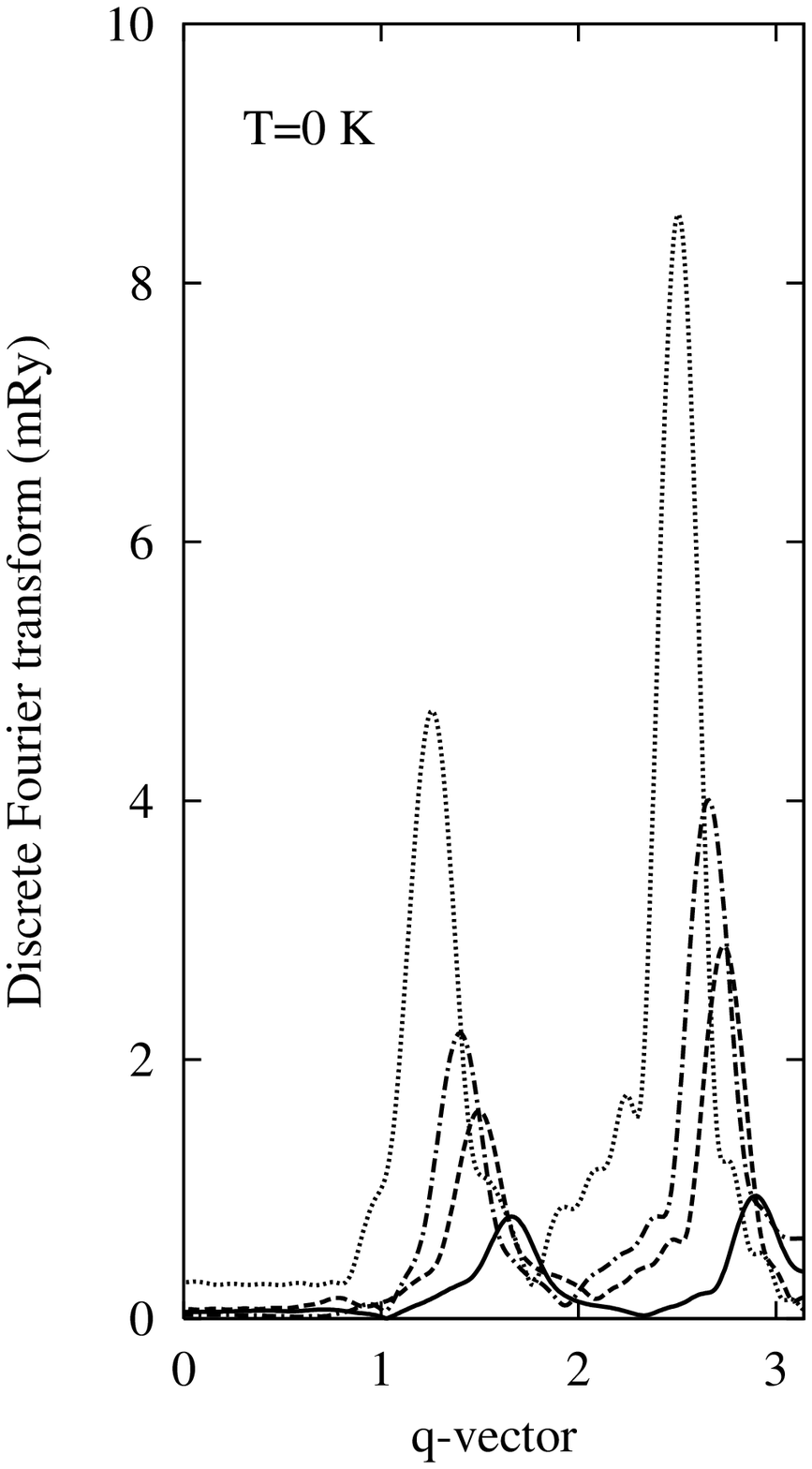}
\includegraphics{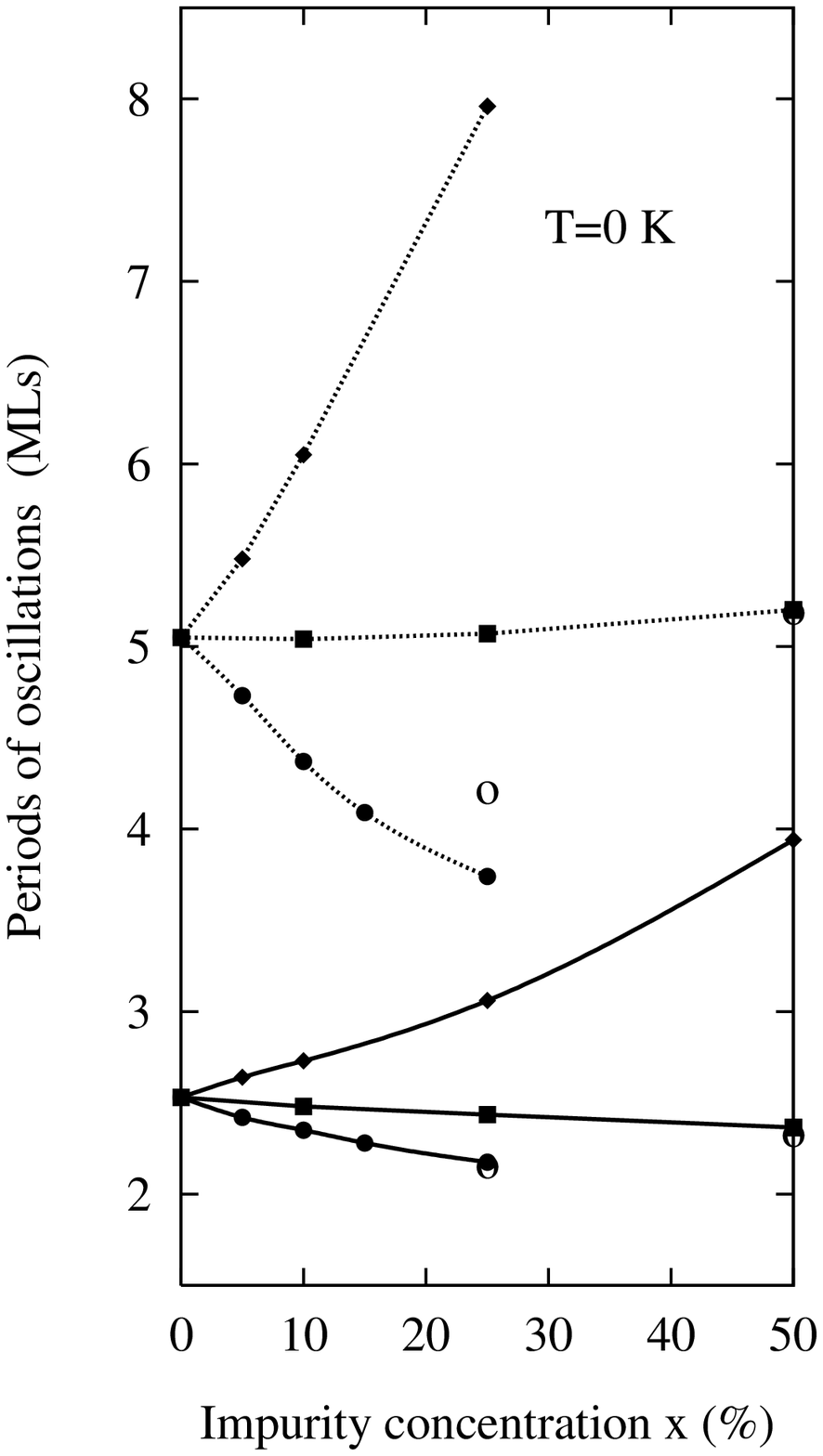}
\vspace*{13cm}
\caption{Left panel: composition dependence of the absolute values of a discrete 
Fourier transform of ${\cal E}_{x}$ at T=0 K for two Co(001) slabs 
each five monolayers thick separated by an fcc-Cu$_{1-x}$Ni$_{x}$ 
alloy spacer: (i) Cu$_{0.75}$Ni$_{0.25}$ (full line), 
(ii) Cu$_{0.85}$Ni$_{0.15}$ (dashed line), Cu$_{0.9}$Ni$_{0.1}$ 
(dashed-dotted line), and (iv) an ideal Cu spacer (dotted line);
from Ref.~\protect\cite{ Kudrnovsky1996b}.
Right panel: composition dependence of the coupling periods 
for two Co(001) slabs each five AL thick separated by an
fcc-Cu$_{1-x}$M$_{x}$ alloy spacer: (i) M=Ni (bullets), (ii) M=Au 
(squares), and (iii) M=Zn (diamonds).
The lines serves as a guide for the eye and distinguish between short 
(full lines) and long (dotted lines) period oscillations. 
Open circles for Cu$_{0.75}$Ni$_{0.25}$ and Cu$_{0.5}$Au$_{0.5}$ 
represent the approximate virtual-crystal values. The periods are 
given in ALs; from Ref.~\protect\cite{ Kudrnovsky1996b}.}
\label{ fig_alloys}
\end{figure}

\subsection{Numerical studies of IEC in presence of alloy disorder}

The interlayer exchange coupling through a Cu$_{1-x}$Ni$_x$ spacer layer 
has been studied experimentally by several groups \cite{ Okuno1993, 
Parkin1993, Bobo1993}. These authors have observed a decrease of oscillation 
periods with increasing Ni concentration, which they have attributed to the 
shrinking of the Fermi surface. 

First-principles calculations of IEC for the Co/Cu$_{1-x}$Ni$_x$/Co(001),
 Co/Cu$_{1-x}$Zn$_x$/Co(001), and Co/Cu$_{1-x}$Au$_x$/Co(001) systems 
based upon the ``vertex cancellation'' theorem have been performed by 
Kudrnovsk\'y {\em et al.\/} \cite{ Kudrnovsky1996b}. Their results are 
shown in Fig.~\ref{ fig_alloys}. The systematic shift of the oscillation periods 
with impurity concentration appears clearly. The decrease (resp. increase) 
of oscillation periods with increasing concentration of Ni (resp. Zn) is 
clearly due to decrease (resp. increase) number of conduction electrons. 
Alloying with an isoelectronic metal (Au) on the other, yields quasi-constant
oscillation periods.


\begin{references}
\bibitem[*]{e-mail} Electronic address: {\tt bruno@mpi-halle.de}

\bibitem{ Parkin1990} S.S.P.~Parkin, N.~More, and K.P.~Roche, 
Phys. Rev. Lett. {\bf 64}, 2304 (1990).

\bibitem{ Yafet1987} Y.~Yafet, Phys. Rev. B {\bf 36}, 3948 (1987).

\bibitem{ Chappert1991} C.~Chappert and J.P.~Renard, Europhys. Lett.
{\bf 15}, 553 (1991).

\bibitem{ Bruno1991} P.~Bruno and C.~Chappert, Phys. Rev. Lett. {\bf
67}, 1602 (1991); {\bf 67}, 2592(E) (1991).

\bibitem{ Bruno1992} Phys. Rev. B {\bf
46}, 261 (1992).

\bibitem{ Coehoorn1991} R.~Coehoorn, Phys. Rev. B {\bf 44}, 9331
(1991).

\bibitem{ Barnas1992} J.~Barna\' s, J. Magn. Magn. Mater. {\bf 111},
L215 (1992).

\bibitem{ Erickson1993} R.P.~Erickson, K.B.~Hathaway, and J.R.~Cullen,
Phys. Rev. B {\bf 47}, 2626 (1993).

\bibitem{ Slonczewski1993} J.C.~Slonczewski, J. Magn. Magn. Mater. {\bf
126}, 374 (1993).

\bibitem{ Edwards1991} D.M.~Edwards, J.~Mathon, R.B.~Muniz, and
M.S.~Phan, Phys. Rev. Lett. {\bf 67}, 493 (1991).

\bibitem{ Mathon1992} J.~Mathon,
M.~Villeret, and D.M.~Edwards, J. Phys. Condens. Mat. {\bf 4},
9873 (1992).

\bibitem{ Wang1990} Y.~Wang, P.M.~Levy, and J.L.~Fry, Phys. Rev.
Lett. {\bf 65}, 2732 (1990). 

\bibitem{ Shi1992} Z.P.~Shi, P.M.~Levy, and J.L.~Fry,
Phys. Rev. Lett. {\bf 69}, 3678 (1992).

\bibitem{ Bruno1992b} P.~Bruno, J. Magn. Magn. Mater. {\bf 116}, L13
(1992).

\bibitem{ Herman1991} F.~Herman, J.~Sticht, and M.~van Schilfgaarde,
J. Appl. Phys. {\bf 69}, 4783 (1991); in {\em Magnetic Thin
Films, Multilayers and Surfaces\/}, edited by S.S.P.~Parkin,
H.~Hopster, J.P.~Renard, T.~Shinjo, and W.~Zinn, Symposia Proceedings No. 
231 (Materials Research Society, Pittsburg, 1992).

\bibitem{ Krompiewski1993} S.~Krompiewski, U.~Krey, and J.~Pirnay, J.
Magn. Magn. Mater. {\bf 121}, 238 (1993).

\bibitem{ Krompiewski1994} S.~Krompiewski, F.~S\"uss, and U.~Krey, 
Europhys. Lett. {\bf 26}, 303 (1994).

\bibitem{ Schilfgaarde1993} M.~van Schilfgaarde and F.~Herman, Phys. Rev. Lett. 
{\bf 71}, 1923 (1993).

\bibitem{ Lang1993} P.~Lang, L.~Nordstr\"om, R.~Zeller, and P.H.~Dederichs, 
Phys. Rev. Lett. {\bf 71}, 1927 (1993).

\bibitem{ Nordstrom1994} L.~Norstr\"om, P.~Lang, R.~Zeller, and P.H.~Dederichs, 
Phys. Rev. B {\bf 50}, 13058 (1994).

\bibitem{ Kudrnovsky1994} J.~Kudrnovsk\'y, V.~Drchal, I.~Turek, and 
P.~Weinberger, Phys. Rev. B {\bf 50}, 16105 (1994).

\bibitem{ Lee1995} B.~Lee and Y.-C.~Chang, Phys. Rev. B {\bf 52}, 3499 (1995). 

\bibitem{ Kudrnovsky1996} J.~Kudrnovsk\'y, V.~Drchal, I.~Turek, M.~\v{S}ob, and 
P.~Weinberger, Phys. Rev. B {\bf 53}, 5125 (1996).

\bibitem{ Lang1996} P.~Lang, L.~Nordstr\"om, K.~Wildberger, R.~Zeller, 
P.H.~Dederichs, and T.~Hoshino, Phys. Rev. B {\bf 53}, 9092 (1996).

\bibitem{ Drchal1996} V.~Drchal, J.~Kudrnovsk\'y, I.~Turek, and P.~Weinberger,
Phys. Rev. B {\bf 53}, 15036 (1996).

\bibitem{ Stiles1996} M.D.~Stiles, J. Appl. Phys. {\bf 79}, 5805 (1996).

\bibitem{ Albuquerque1996} J.~d'Albuquerque e Castro, J.~Mathon, M.~Villeret, 
and A.~Umerski, Phys. Rev. B {\bf 53}, R13306 (1996). 

\bibitem{ Lee1996} B.~Lee and Y.C.~Chang, Phys. Rev. B {\bf 54}, 13034 (1996).

\bibitem{ Mathon1997} J.~Mathon, M.~Villeret, A.~Umerski, R.B.~Muniz, 
J.~d'Albuquerque e Castro, and D.M.~Edwards, Phys. Rev. B {\bf 56}, 11797 
(1997).

\bibitem{ Costa1997} A.T.~Costa, Jr., J.~d'Albuquerque e Castro, and R.B.~Muniz, 
Phys. Rev. B {\bf 56}, 13697 (1997).

\bibitem{ Bruno1993} P.~Bruno, J. Magn. Magn. Mater. {\bf 121}, 248
(1993).

\bibitem{ Bruno1995} P.~Bruno, Phys. Rev. B {\bf 52}, 411 (1995).

\bibitem{ Stiles1993} M.D.~Stiles, Phys. rev. B {\bf 48} 7238 (1193).

\bibitem{ Loly1983} P.D.~Loly and J.B.~Pendry, J. Phys. C: 
Solid State Phys. {\bf 16}, 423 (1983).

\bibitem{ Wachs1986} A.L.~Wachs, A.P.~Shapiro, T.C.~Hsieh, and T.-C.~Chiang, 
Phys. Rev. B {\bf 33}, 1460 (1986). 

\bibitem{ Lindgren1987} S.\AA\ Lindgren and L.~Walld\'en, Phys. Rev. Lett. 
{\bf 59}, 3003 (1987).

\bibitem{ Lindgren1988} S.\AA\ Lindgren and L.~Walld\'en, Phys. Rev. Lett. 
{\bf 61}, 2894 (1988).

\bibitem{ Lindgren1988b} S.\AA\ Lindgren and L.~Walld\'en, Phys. Rev. B 
{\bf 38}, 3060 (1988).

\bibitem{ Lindgren1989} S.\AA\ Lindgren and L.~Walld\'en, J. Phys.: Condens. 
Matter {\bf 1}, 2151 (1989).

\bibitem{ Miller1988} T.~Miller, A.~Samsavar, G.E.~Franklin, and T.-C.~Chiang, 
Phys. Rev. Lett. {\bf 61}, 1404 (1988).

\bibitem{ Mueller1989} M.A.~Mueller, A.~Samsavar, T.~Miller, and T.-C.~Chiang, 
Phys. Rev. B {\bf 40}, 5845 (1989).

\bibitem{ Mueller1990} M.A.~Mueller, T.~Miller, and T.-C.~Chiang, 
Phys. Rev. B {\bf 41}, 5214 (1990).

\bibitem{ Jalochowski1992} M.~Ja\l ochowski, E.~Bauer, H.~Knoppe, and 
G.~Lilienkamp, Phys. Rev. B {\bf 45}, 13607 (1992). 

\bibitem{ Brookes1991} N.B.~Brookes, Y.~Chang, and P.D.~Johnson, Phys. Rev. 
Lett. {\bf 67}, 354 (1991).

\bibitem{ Ortega1992} J.E.~Ortega and F.J.~Himpsel, Phys. Rev. Lett. {\bf 69}, 
844 1(992).

\bibitem{ Ortega1993} J.E.~Ortega, F.J.~Himpsel, G.J.~Mankey, and R.F.~Willis, 
Phys. Rev. B {\bf 47}, 1540 (1993).

\bibitem{ Ortega1993b} J.E.~Ortega, F.J.~Himpsel, G.J.~Mankey, and R.F.~Willis, 
J. Appl. Phys. {\bf 73}, 5771 (1993).

\bibitem{ Garrison1993} K.~Garrison, Y.~Chang, and P.D.~Johnson, Phys. Rev. 
Lett. {\bf 71}, 2801 (1993).

\bibitem{ Carbone1993} C.~Carbone, E.~Vescovo, O.~Rader, W.~Gudat, and 
W.~Eberhardt, Phys. Rev. Lett. {\bf 71}, 2805 (1993).

\bibitem{ Smith1994} N.V.~Smith, N.B.~Brookes, Y.~Chang, and P.D.~Johnson, 
Phys. Rev. B {\bf 49}, 332 (1994).

\bibitem{ Johnson1994} P.D.~Johnson, K.~Garrison, Q.~Dong, N.V.~Smith, D.~Li, 
J.~Mattson, J.~Pearson, and S.D.~Bader, Phys. Rev. B {\bf 50}, 8954 (1994).

\bibitem{ Himpsel1995} F.J.~Himpsel and O.~Rader, Appl. Phys. Lett. {\bf 67}, 
1151 (1995).

\bibitem{ Crampin1996} S.~Crampin, S.~De Rossi, and F.~Ciccaci, Phys. Rev. B 
{\bf 53}, 13817 (1996).

\bibitem{ Segovia1996} P.~Segovia, E.G.~Michel, and J.E.~Ortega, Phys. Rev. 
Lett. {\bf 77}, 3455 (1996).

\bibitem{ Klaesges1998} R.~Kl\"asge, D.~Schmitz, C.~Carbone, W.~Eberhardt, 
P.~Lang, R.~Zeller, and P.H.~Dederichs, Phys. Rev. B {\bf 57}, R696 (1998).

\bibitem{ Himpsel1991} F.J.~Himpsel, Phys. Rev. B {\bf 44}, 5966 (1991).

\bibitem{ Ortega1993c} J.E.~Ortega and F.J.~Himpsel, Phys. Rev. B {\bf 47}, 
16441 (1993).

\bibitem{ Li1995} D.~Li, J.~Pearson, J.E.~Mattson, S.D.~Bader, and P.D.~Johnson, 
Phys. Rev. B {\bf 51}, 7195 (1995).

\bibitem{ Himpsel1995b} F.J.~Himpsel, J. Electr. Microsc. and Rel. Phenom. 
{\bf 75}, 187 (1995).

\bibitem{ Li1997} D.~Li, J.~Pearson, S.D.~Bader, E.~Vescovo, D.-J.~Huang, 
P.D.~Johnson, and B.~Heinreich, Phys. Rev. Lett. {\bf 78}, 1154 (1997).

\bibitem{ Koike1994} K.~Koike, T.~Furukawa, G.P.~Cameron, and Y.~Murayama, 
Phys. Rev. B {\bf 50}, 4816 (1994).

\bibitem{ Furukawa1996} T.~Furukawa and K.~Koike, Phys. Rev. B {\bf 54}, 17896 
(1996).

\bibitem{ Bennett1990} W.R.~Bennett, W.~Schwarzacher, and W.F.~Egelhoff, Jr., 
Phys. Rev. Lett. {\bf 65}, 3169 (1990).

\bibitem{ Katayama1993} T.~Katayama, Y.~Suzuki, M.~Hayashi, and A.~Thiaville, 
J. Magn. Magn. Mater. {\bf 126}, 527 (1993).

\bibitem{ Carl1995} A.~Carl and D.~Weller, Phys. Rev. Lett. {\bf 74}, 190 
(1995).

\bibitem{ Megy1995} R.~M\'egy, A.~Bounouh, Y.~Suzuki, P.~Beauvillain, P.~Bruno, 
C.~Chappert, B.~L\'ecuyer, and P.~Veillet, Phys. Rev. B {\bf 51}, 5586 (1995).

\bibitem{ Bruno1996c} P.~Bruno, Y.~Suzuki and C.~Chappert, Phys. Rev. B 
{\bf 53}, 9214 (1996).

\bibitem{ Suzuki1998} Y.~Suzuki, T.~Katayama, P.~Bruno, S.~Yuasa, and 
E.~Tamura, Phys. Rev. Lett. {\bf 80}, 5200 (1998).

\bibitem{ Luce1996} T.A.~Luce, W.~H\"ubner, and K.H.~Bennemann, Phys. Rev. Lett. 
{\bf 77}, 2810 (1996).

\bibitem{ Kirilyuk1996} A.~Kirilyuk, Th.~Rasing, R.~M\'egy, and P.~Beauvillain, 
Phys. Rev. Lett. {\bf 77}, 4608 (1996).

\bibitem{ Weber1996} W.~Weber, A.~Bischof, R.~Allenspach, C.~W\"ursch, 
C.H.~Back, and D.~Pescia, Phys. Rev. Lett. {\bf 76}, 3424 (1996).

\bibitem{ Back1997} C.H.~Back, W.~Weber, C.~W\"ursch, A.~Bischof, D.~Pescia, 
and R.~Allespach, J. Appl. Phys. {\bf 81}, 5054 (1997).

\bibitem{ Halse1969} M.R.~Halse, Philos. Trans. R. Soc. London A {\bf 265}, 
507 (1969).

\bibitem{ Johnson1992} M.T.~Johnson, S.T.~Purcell, N.W.E.~McGee, R.~Coehoorn, 
J.~aan de Stegge, and W.~Hoving, Phys. Rev. Lett. {\bf 68}, 2688 (1992).

\bibitem{ Weber1995} W.~Weber, R.~Allenspach, and A.~Bischof, Europhys. Lett. 
{\bf 31}, 491 (1995).

\bibitem{ Unguris1993} J.~Unguris, R.J.~Celotta, and D.T.~Pierce, J. Magn. 
Magn. Mater. {\bf 127}, 205 (1993).

\bibitem{ Fuss1992} A.~Fuss, S.~Demokritov, P.~Gr\"unberg, and W.~Zinn, 
J. Magn. Magn. Mater. {\bf 103}, L221 (1992).

\bibitem{ Unguris1994} J.~Unguris, R.J.~Celotta, and D.T.~Pierce, J. Appl. Phys.
{\bf 75}, 6437 (1994).

\bibitem{ Unguris1997} J.~Unguris, R.J.~Celotta, and D.T.~Pierce, Phys. Rev. 
Lett. {\bf 79}, 2734 (1998).

\bibitem{ Parkin1991} S.S.P.~Parkin, R.~Bhadra, and K.P.~Roche, Phys. Rev. Lett. 
{\bf 66}, 2152 (1991).

\bibitem{ Mosca1991} D.H.~Mosca, F.~P\'etroff, A.~Fert, P.A.~Schroeder, 
W.P.~Pratt, Jr., R.~Laloee, and S.~Lequien, J. Magn. Magn. Mater. {\bf 94}, L1 
(1991).

\bibitem{ Petroff1991} F.~P\'etroff, A.~Barth\'e\'emy, D.H.~Mosca, D.K.~Lottis, 
A.~Fert, P.A.~Schroeder, W.P.~Pratt, Jr., R.~Laloee, and S.~Lequien, Phys. Rev. 
B {\bf 44}, 5355 (1991).

\bibitem{ Miguel1991} J.J.~de Miguel, A.~Cebollada, J.M.~Gallego, R.~Miranda, 
C.M.~Schneider, P.~Schuster, and J.~Kirschner, J. Magn. Magn. Mater. {\bf 93}, 
1 (1991).

\bibitem{ Okuno1993} S.N.~Okuno and K.~Inomata, Phys. Rev. Lett. {\bf 70}, 
1771 (1993).

\bibitem{ Parkin1993} S.S.P.~Parkin, C.~Chappert, and F.~Herman, Europhys. 
Lett. {\bf 24}, 71 (1993).

\bibitem{ Bobo1993} J.-F.~Bobo, L.~Hennet, and M.~Pi\'ecuch, Europhys. Lett.
{\bf  24}, 139 (1993).

\bibitem{ Bruno1998} P.~Bruno, Los Alamos Preprint Server, http://xxx.lanl.gov 
cond-mat/9808091 

\bibitem{ Bruno1993b} P.~Bruno, Europhys. Lett. {\bf 23}, 615 (1993).

\bibitem{ Bloemen1994} P.J.H.~Bloemen, M.T.~Johnson, M.T.H.~van de Vorst, 
R.~Coehoorn, J.J.~de Vries, R.~Jungblut, J.~aan de Stegge, A.~Reiders, and 
W.J.M.~de Jonge, Phys. Rev. Lett. {\bf 72}, 764 (1994).

\bibitem{ Okuno1994} S.N.~Okuno and K.~Inomata, Phys. Rev. Lett. {\bf 72}, 
1553 (1994).

\bibitem{ Vries1995} J.J.~de Vries, A.A.P.~Schudelaro, R.~Jungblut, 
P.J.H.~Bloemen, A.~Reinders, J.~Kohlhepp, R.~Coehoorn, and de W.J.M.~Jonge, 
Phys. Rev. Lett. {\bf 75}, 1306 (1995).

\bibitem{ Okuno1995} S.N.~Okuno and K.~Inomata, J. Phys. Soc. Japan {\bf 64}, 
3631 (1995).

\bibitem{ Bounouh1996} A.~Bounouh, P.~Beauvillain, P.~Bruno, C.~Chappert, 
R.~M\'egy, and P.~Veillet, Europhys. Lett. {\bf 33}, 315 (1996).

\bibitem{ Bruno1996} P.~Bruno, J. Magn. Magn. Mater. {\bf 164}, 27 (1996).

\bibitem{ Kudrnovsky1997} J.~Kudrnovsk\'y, V.~Drchal, P.~Bruno, I.~Turek, and 
P.~Weinberger, Phys. Rev. B. {\bf 56}, 8919 (1997).

\bibitem{ Hohenberg1964} P.~Hohenberg and W.~Kohn, Phys. Rev. {\bf 136}, 
B864 (1964).

\bibitem{ Kohn1965} W.~Kohn and L.J.~Sham, Phys. Rev. {\bf 140}, A1133 (1965).

\bibitem{ Barth1972} U.~von Barth und L.~Hedin, J. Phys. C: Solid State 
Phys. {\bf 5}, 1629 (1972).

\bibitem{ Dederichs1984} P.H.~Dederichs, S.~Bl\"ugel, R.~Zeller, and H.~Akai, 
Phys. Rev. Lett. {\bf 53}, 2512 (1984).

\bibitem{ Mackintosh1980} A.R.~Mackintosh and O.K.~Andersen, in {\em
Electrons at the Fermi Surface\/}, edited by M.~Springford
(Cambridge University Press, Cambridge, England, 1980), p. 149.

\bibitem{ Weinert1985} M.~Weinert, R.E.~Watson, and J.W.~Davenport, 
Phys. Rev. B {\bf 32}, 2115 (1985).

\bibitem{ Harris1985} J.~Harris, Phys. Rev. B {\bf 31}, 1770 (1985).

\bibitem{ Foulkes1989} W.M.C.~Foulkes and R.~Haydock, Phys. Rev. B {\bf 39}, 
12520 (1989).

\bibitem{ Read1989} A.J.~Read and R.J.~Needs, J. Phys.: Condens Matter {\bf 1}, 
7565 (1989).

\bibitem{ Finnis1990} M.W.~Finnis, J. Phys.: Condens. Matter {\bf 2} 331, (1990).

\bibitem{ Robertson1991} I.J.~Robertson and B.~Farid, Phys. Rev. Lett. {\bf 66}, 
3265 (1991).

\bibitem{ Kudrnovsky_PC} J.~Kudrnovsk\'y, V.~Drchal, and I.~Turek, private 
communication.

\bibitem{ Bruno1996b} P.~Bruno, J.~Kudrnovsk\'y, V.~Drchal, and I.~Turek, 
Phys. Rev. Lett. {\bf 76}, 4253 (1996).

\bibitem{ Ducastelle1975} F.~Ducastelle, J. Phys. C: Solid State Phys. {\bf 8},
3297 (1975).

\bibitem{ Kudrnovsky1996b} J.~Kudrnovsk\'y, V.~Drchal, P.~Bruno, I.~Turek, 
and P.Weinberger, Phys. Rev. B {\bf 54}, 3738 (1996).


\end{references}
\end{document}